\documentclass[journal,twoside,web]{ieeecolor}

    \usepackage[pdftex]{graphicx}
    \DeclareGraphicsExtensions{.pdf,.jpeg,.png}

\usepackage{generic}
\usepackage{gensymb}
\usepackage{epstopdf}
\usepackage{graphicx}
\usepackage{amssymb}
\usepackage{amsmath}
\usepackage{float}
\usepackage{amsfonts, bm}
\usepackage{color}
\usepackage[noadjust]{cite}

\newcommand{\rv}[1]{\bm{#1}}

\DeclareMathOperator{\Ex}{\mathcal{E}}
\DeclareMathOperator{\sinc}{sn}
\DeclareMathOperator*{\sol}{sol}
\DeclareMathOperator{\trace}{tr}

\DeclareMathOperator{\diag}{diag}
\DeclareMathOperator{\wpone}{wp1}
\DeclareMathOperator{\as}{a.s.}

\renewcommand{\theenumi}{\alph{enumi}}

\newtheorem{assm}{Assumption}
\newtheorem{th1}{Theorem}
\newtheorem{lm1}{Lemma}

\newtheorem{rem1}{Remark}

\title{
Causal, Stochastic MPC for Wave Energy Converters
}

\author{Connor H. Ligeikis and Jeffrey T. Scruggs
\thanks{The first author was supported by an NSF Graduate Research Fellowship. This funding is gratefully acknowledged. Views expressed in this paper are those of the authors and do not necessarily reflect those of the National Science Foundation.}
\thanks{C. Ligeikis and J. Scruggs are with the Department of Civil \& Environmental Engineering,
      	University of Michigan, Ann Arbor, MI, 48109.
      	Phone: \mbox{734-764-1812}, email: \mbox{jscruggs@umich.edu}, \mbox{ligeikis@umich.edu}}
}

\begin{document}

\maketitle

\begin{abstract}                
We implement a causal model predictive control (MPC) strategy to maximize power generation from a wave energy converter (WEC) system, for which the power take-off (PTO) systems have both hard stroke (i.e., displacement) limits and force ratings. 
The approach models the WEC dynamics in discrete-time, in a manner that exactly preserves energy-flow quantities, and assumes a stationary stochastic disturbance model for the incident wave force. 
The control objective is to maximize the expected power generation in stationarity, while accounting for parasitic losses in the power train.
PTO stroke measurements are assumed to be available for real-time feedback, as well as the free-surface elevation of the waves at a designated location relative to the WEC, and the open-loop dynamics of the WEC are assumed to be linear and time-invariant.
Mean-square stability of the MPC algorithm is proven.
The methodology is illustrated in a simulation example pertaining to a heaving cylindrical buoy.
\end{abstract}

\begin{IEEEkeywords}
Energy Systems, Model Predictive Control, Passivity
\end{IEEEkeywords}

\section{Introduction}

Wave energy converters have been researched extensively for many decades \cite{mccormick2013ocean}. 
The vast array of technologies proposed to harness the resource is diverse and defies easy generalization \cite{antonio2010wave,clemente2021potential,guo2021review}.
However, most WEC devices are comprised of a mechanical system, either floating at the ocean surface or mounted to the ocean floor, which is dynamically-excited by propagating waves \cite{falnes2007review}. 
This mechanical motion is coupled to one or more PTOs (which can be either electromechanical or hydraulic) which extract power from the structural dynamics \cite{salter2002power,jusoh2019hydraulic,ahamed2020advancements,ahamed2022review}.
The PTOs are interfaced with a power train that transmits extracted power to a localized storage system (such as a electrical supercapacitor, a flywheel, or a hydraulic accumulator) which serves as a buffer, allowing for smoothed power to be transmitted to a utility grid \cite{murray2012supercapacitor,tedeschi2013modeling}. 
To maximize power production in irregular (i.e., stochastic) waves, there is a tangible advantage to the control of the power extracted by the PTOs, based on real-time feedback measurements of the WEC system's dynamic response \cite{hals2011comparison,scruggs2013optimal,ringwood2014energy,coe2017comparison}. 
The synthesis of optimal controllers for WEC systems is a topic that continues to be an area of active research.

Early studies on the optimal control of WECs focused primarily on linear feedback laws.
Beginning with the seminal papers by Falnes \cite{falnes1980radiation} and Evans \cite{evans1981power}, it was recognized that in monochromatic waves, power is maximized by imposing an effective impedance relationship between the PTO velocity vector, and the opposing force or torque vector, with the optimal impedance matrix equal to the Hermitian adjoint of the driving point impedance of the WEC system.
This result is equivalent to the impedance matching technique used in antenna arrays and other electromagnetic technologies. 
Impedance matching is of limited practicality for at least three distinct reasons.
The first is that, when extended to the case of stochastic waves, the optimal impedance-matching feedback law is anti-causal, and therefore can only be implemented if an accurate forecast is available for the incident wave forces on the WEC system \cite{nebel1992maximizing,zhan2018linear}.
Secondly, the technique presumes that the PTOs have no constraints.
This is especially problematic because the dynamic response of the controlled WEC system often deviates further from equilibrium than the uncontrolled system \cite{evans1981maximum,hals2011constrained,bacelli2013constrained}.
As such, it is important to account for constraints on the PTOs, particularly their limits on displacement, and their force ratings. 
Thirdly, impedance matching does not extend easily to cases in which the WEC system exhibits nonlinear dynamics \cite{penalba2017mathematical}. 

For all these reasons, the last decade has seen a rally around the use of MPC techniques to maximize the energy generated by WEC systems \cite{cretel2010application,%
hals2011constrained,%
richter2012nonlinear,%
bacelli2013constrained,%
li2014model,%
tom2015experimental,%
faedo2017optimal,%
zhong2018efficient}.
MPC is especially well-suited to address the second problem discussed above, concerning the accommodation of PTO constraints.
Additionally, many nonlinear MPC techniques exist which can readily be applied to WEC systems with nonlinear plant dynamics \cite{allgower2012nonlinear,li2017nonlinear,na2018nonlinear,haider2021real}.
However, most of the MPC techniques implemented for WEC systems do not directly address the first problem, related to causality.
Indeed, many MPC techniques for WECs assume an accurate real-time prediction exists over a long receding horizon, for the incident wave force.
Predicated on this assumption, MPC is implemented at each discrete time step by optimizing the control force/torque trajectory for the PTOs over this receding horizon.
The first component of this optimized trajectory is then implemented by the controller.
At the next time step, the receding horizon for the wave force forecast is advanced, the PTO input trajectory is re-optimized, and the process is thus repeated ad infinitum. 

To implement the MPC technique described above, it is necessary to somehow forecast the incident wave force, and many techniques for this have been proposed in the literature.
Some obtain the forecast by assuming that measurements of free-surface elevations are available at a distance up-wave from the WEC
\cite{previsic2021ocean}.
However, most approaches presume that the incident wave force on the WEC can be measured \cite{abdelhalik2017}, and predict the future force trajectories from past data using one of several algorithms, including fully-empirical curve-fitting algorithms as well as model-based algorithms such as the extended Kalman filter \cite{fusco2010short,%
schoen2011wave,%
li2012wave,%
korde2015near,%
davis2020wave}.

The primary purpose of this paper is to provide an alternative to the above prediction-based techniques, and to establish that WEC MPC algorithms need not include an explicit wave force forecast at all, to achieve near-optimal performance.
Specific contributions of the paper are:
\begin{enumerate}
\item We illustrate an entirely causal technique for implementing MPC in WEC applications, which only requires that the power spectral density of the sea state be known. 
\item We illustrate that if free surface elevations are measured and available for feedback, these measurements may be systematically incorporated into the MPC algorithm.
\item We illustrate how parasitic losses in the PTO and power train can be systematically incorporated into the MPC optimization objective.
\item We prove that the proposed MPC algorithm is stable in the sense of Lyapunov (for free response), in the bounded-input bounded-state sense (for transient disturbances), and in the mean-square sense (for stationary stochastic disturbances). 
\end{enumerate}
The scope of the paper is limited to WEC systems with linear plant dynamics, but many of the techniques we discuss may be extended to the nonlinear case.  

The paper is organized as follows.
Section II establishes the discrete-time stochastic modeling framework. 
Section III formalizes the WEC feedback control problem, and establishes the MPC control framework. 
Section IV focuses on the trajectory optimization problem for the MPC controller, illustrates how this algorithm can be implemented as a convex optimization, and provides the stability results. 
Section V provides a numerical example of the implementation of the MPC algorithm for a heaving, cylindrical WEC.
Finally, Section VI draws some conclusions. 

\subsection{Notation and terminology}

Sets are denoted in blackboard font, e.g., $\mathbb{R}$, $\mathbb{C}$, $\mathbb{Z}$, and so on.
We notate $\mathbb{R}_{\geqslant 0}$ and $\mathbb{Z}_{\geqslant 0}$ as the sets of all nonnegative reals and integers, respectively.
The notation $\mathbb{C}_{\leqslant 1}$ (and $\mathbb{C}_{<1}$) denote the complex numbers with moduli less than (strictly less than) $1$. Similar definitions hold for the opposite inequalities. 
For a matrix $Q \in \mathbb{R}^{n\times m}$ or $\mathbb{C}^{n\times m}$, $Q^T$ and $Q^H$ are the transpose and Hermitian transpose, respectively. 
For a vector $q \in \mathbb{R}^n$ or $\mathbb{C}^n$ and $p \in \mathbb{Z}_{\geqslant 1}$, the norm $\|q\|_p = [ \sum_{i=1}^n |q_i|^p ]^{1/p}$ and the norm $\|q\|_\infty = \max_i |q_i|$. 
For a matrix $Q \in \mathbb{R}^{n\times m}$ or $\mathbb{C}^{n\times m}$ and $p \in \mathbb{Z}_{\geqslant 1} \cup \{\infty\}$, the norm $\|Q\|_p$ is the induced norm, i.e., $\sup_{\|q\|_p=1} \|Qq\|_p$.
When no subscript is specified for a norm, i.e., $\| \cdot \|$, $p = 2$ is assumed.
For a matrix $Q=Q^H\geqslant 0$ and a compatible vector $q$, we denote $\| q \|_Q = (q^HQq)^{1/2}$.

For $p$ finite, the set $\mathbb{L}_p$ is the set of all Lebesgue-integrable functions $f(x)$ of a real variable $x$, such that $\|f\|_{\mathbb{L}_p} \triangleq ( \int_{-\infty}^\infty \| f(x) \|^p dx )^{1/p} < \infty$. The set $\mathbb{L}_\infty$ is the set of all Lebesgue-integrable functions $f(x)$ such that $\|f\|_{\mathbb{L}_\infty} \triangleq \sup_{x\in\mathbb{R}} \| f(x) \| < \infty$. 
For a continuous-time function $f(t)$ we denote its Fourier transform as $\hat{f}(j\omega)$.
For a discrete-time function $f_k$, we denote its $z$-transform as $\bar{f}(z)$. 
The sets $\mathbb{H}_2$ and $\mathbb{H}_\infty$ are the Hardy spaces, comprised of complex functions $\bar{f}(z)$ which are analytic for all $z \in \mathbb{C}_{>1}$, and which satisfy an appropriate norm on the boundary. For $\bar{f} \in \mathbb{H}_2$, we requires that $\|\bar{f}\|_{\mathbb{H}_2} \triangleq (\tfrac{1}{2\pi}\int_{-\pi}^\pi \| \bar{f}(e^{j\Omega} \|^2 d\Omega )^{1/2} < \infty$. 
For $\bar{f} \in \mathbb{H}_\infty$ we require that $\|\bar{f}\|_{\mathbb{H}_\infty} \triangleq \sup_{\Omega \in [-\pi,\pi]} \| \bar{f}(e^{j\Omega}) \|^2 < \infty$. 

Stochastic processes and sequences, and other random variables, are denoted in bold. 
For a random variable $\rv{x}$, a particular realization is denoted by the same character in italics, i.e., $x$. 
For a random variable $\rv{x}$ and information $\mathbb{I}$, we denote $\mathcal{E}\{\rv{x} | \mathbb{I}\}$ as the conditional expectation of $\rv{x}$.
For two random variables $\rv{x}$ and $\rv{y}$, the expression $\mathcal{E}\{ \rv{y} | \rv{x} \}$ is $\rv{y}$ conditioned on $\rv{x}$, and should be interpreted as a function of random variable $\rv{x}$.  
With reference to a stochastic sequence $\{ \rv{x}_k : k \in \mathbb{Z} \}$ we refer to a particular realization of a component or sub-sequence as \emph{data}. 
The abbreviations $\wpone$ and $\as$ stand for ``with probability $1$'' and ``almost sure.''

\section{System Modeling}

\subsection{Continuous-time model}

For a WEC system with $n_p$ PTOs, let $\rv{d}(t) \in \mathbb{R}^{n_p}$ be the vector of PTO displacements, and let $\rv{v}(t) \triangleq \tfrac{d}{dt}\rv{d}(t)$ be the resultant velocity vector. 
Further, let $\rv{u}(t)\in\mathbb{R}^{n_p}$ be the vector of colocated forces (or torques) associated with the PTOs.
We assume that $\rv{u}(t)$ can be made to track a command with high bandwidth, and therefore can be treated as an input that can be controlled directly.
Let $\rv{f}(t) \in \mathbb{R}^{n_f}$ be the vector of forces on the various mechanical degrees of freedom of the WEC system, due to the incident waves.
Then we presume linear time-invariant (LTI) mappings $Z_{fd}$ and $Z_{ud}$, such that
\begin{equation}
\rv{d}(t) = \left( Z_{fd} \rv{f} \right)(t) + \left( Z_{ud} \rv{u} \right)(t).
\end{equation}
These mappings are uniquely characterized by their frequency response functions $\hat{Z}_{fd}(j\omega)$ and $\hat{Z}_{ud}(j\omega)$, which are obtained from the hydrodynamic analysis of the WEC. 
For simple WEC shapes, they can be obtained via analytical series solutions to the partial differential equations characterizing the fluid-structure interaction \cite{mavrakos1997comparison}.
For more realistic WEC shapes, they can be obtained by via finite-element techniques \cite{folley2012review}.

Let $\rv{a}(t)\in\mathbb{R}$ be the free surface elevation at some fixed location in the ocean, not necessarily colocated with the location of the WEC system.
Then we assume $\rv{a}$ is a stationary stochastic process with known power spectral density (PSD) $S_a(\omega)$ and propagatory direction, with the convention that
\begin{equation}
\mathcal{E}\left\{ \rv{a}(t+\tau) \rv{a}(t) \right\} = \frac{1}{2\pi} \int_{-\infty}^\infty e^{j\omega \tau} S_a(\omega) d\omega
\end{equation}
We assume linear wave theory, and consequently, that there exists a LTI mapping $Z_{af}$ such that 
\begin{equation}
\rv{f}(t) = \left( Z_{af} \rv{a} \right)(t)
\end{equation}
where $Z_{af}$ is uniquely characterized by its frequency-response function $\hat{Z}_{af}(j\omega)$, which must also be obtained through hydrodynamic analysis.
Letting $\rv{d}_f(t) \triangleq (Z_{fd}\rv{f})(t)$, we have that $\rv{d}_f(t)$ is a stationary stochastic process as well.
Defining 
\begin{equation}
\rv{y}_f(t) \triangleq \begin{bmatrix} \rv{d}_f^T(t) & \rv{a}(t) \end{bmatrix}^T
\end{equation}
we have that the joint spectrum of $\rv{y}_f$ is 
\begin{equation}
S_{y_f}(\omega) = \begin{bmatrix} \hat{Z}_{fd}(j\omega) \hat{Z}_{af}(j\omega) \\ 1 \end{bmatrix} S_a(\omega) \begin{bmatrix} \hat{Z}_{fd}(j\omega) \hat{Z}_{af}(j\omega) \\ 1 \end{bmatrix}^H.
\end{equation}
The stochastic system model is then, equivalently,
\begin{equation}
\begin{bmatrix} \rv{d}(t) \\ \rv{a}(t) \end{bmatrix}
=
\rv{y}_f(t) + \begin{bmatrix} I \\ 0 \end{bmatrix} (Z_{ud}\rv{u}) (t) 
\end{equation}
Recalling that $\rv{v}(t) = \tfrac{d}{dt} \rv{d}(t)$, we have that
\begin{equation}
\rv{v}(t) = \tfrac{d}{dt} \rv{d}_f(t) + \left( Z_{uv}\rv{u} \right)(t)
\end{equation}
where mapping $Z_{uv}$ is uniquely characterized by its frequency response functions, as
\begin{align}
\hat{Z}_{uv}(j\omega) =& j\omega \hat{Z}_{ud}(j\omega) \label{Zuv} 
\end{align}
and where we note that the spectrum for $\tfrac{d}{dt}\rv{d}_f$ is $S_{v_f}(\omega) = \omega^2 \hat{Z}_{fd}(j\omega) \hat{Z}_{af}(j\omega) S_a(\omega) \hat{Z}_{fd}^H(j\omega) \hat{Z}_{af}^H(j\omega)$.

\begin{assm} \label{assumption_1}
We assume the following:
\begin{enumerate}
\item 
\label{assumption_1_domains}
$\{ \hat{Z}_{ud}, \hat{Z}_{uv} \} \subset \mathbb{L}_\infty$ and $ \{ S_{y_f}, S_{v_f} \} \subset \mathbb{L}_\infty \cap \mathbb{L}_1$.
\item 
\label{assumption_1_infinity}
The limit $\hat{Z}_{uv}^\infty \triangleq \lim_{\omega\rightarrow\infty} \hat{Z}_{uv}(j\omega)$ exists and is real. 
\item 
\label{assumption_1_L1}
Let $\tilde{Z}_{uv}(t)$ be the impulse response of $Z_{uv}$, i.e., 
\begin{equation}
\tilde{Z}_{uv}(t) = \hat{Z}_{uv}^\infty \delta(t) + \frac{1}{2\pi} \int_{-\infty}^\infty e^{j\omega t} \left( \hat{Z}_{uv}(j\omega) - \hat{Z}_{uv}^\infty \right)d\omega
\end{equation}
with a similar definition for $\tilde{Z}_{ud}(t)$. 
We assume $\{ \tilde{Z}_{uv}, \tilde{Z}_{ud} \} \subset \mathbb{L}_1$.
\item
\label{assumption_1_causality}
$Z_{uv}$ (and therefore $Z_{ud}$) are causal mappings, i.e., $\tilde{Z}_{uv}(t) = \tilde{Z}_{ud}(t) = 0$ for all $t < 0$. 
\item 
\label{assumption_1_passivity}
$Z_{uv}$ is a passive mapping. Equivalently, $\hat{Z}_{uv}$ is positive-real, i.e., 
\begin{equation}\label{passivity_ineq_continuous_time}
\hat{Z}_{uv}(j\omega) + \hat{Z}_{uv}^H(j\omega) \geqslant 0, \ \ \forall \omega \in \mathbb{R} 
\end{equation}
with the inequality holding strictly for all but a countably-finite subset of $\omega \in \mathbb{R}$.
\end{enumerate}
\end{assm}

\begin{rem1}
Assumptions \ref{assumption_1}\ref{assumption_1_domains},   \ref{assumption_1}\ref{assumption_1_infinity}, \ref{assumption_1}\ref{assumption_1_L1}, and \ref{assumption_1}\ref{assumption_1_causality} are stated here explicitly, but are taken for granted in most of the literature on WEC control.
We note that Assumption \ref{assumption_1}\ref{assumption_1_L1} implies that the mappings $Z_{uv}$ and $Z_{ud}$ have bounded $\mathbb{L}_\infty$ gain.
Because of \eqref{Zuv}, Assumption \ref{assumption_1}\ref{assumption_1_domains} necessarily implies that $\hat{Z}_{uv}(0) = 0$, and also that $\hat{Z}_{ud}$ must be strictly proper. 
\end{rem1}
\begin{rem1}
If Assumption \ref{assumption_1}\ref{assumption_1_passivity} is not satisfied, this implies that the WEC possesses an internal energy source. In that case, even in the absence of a disturbance (i.e., with $\rv{f} = 0$), energy can be generated, implying the WEC is a perpetual-motion machine. 
The caveat that \eqref{passivity_ineq_continuous_time} is strict for all but countably-finite frequencies is mild. At these frequencies, it is possible to excite the WEC with a sinusoidal force $\rv{u}(t)$ that results in zero mechanical power injection. For almost all applications, the only (finite) frequency where this can occur is $\omega = 0$. 
\end{rem1}

The power generated by the WEC at time $t$ is denoted $\rv{\varrho}(t)$, and is assumed to be of the form
\begin{equation}\label{pg}
\rv{\varrho}(t) = -\rv{u}^T(t) \rv{v}(t) - \rv{u}^T(t) R \rv{u}(t) - v_d^T \left| \rv{u}(t) \right|
\end{equation}
where $R \in \mathbb{R}^{n_p\times n_p}$ with $R = R^T > 0$, and $v_d \in \mathbb{R}_{\geqslant 0}^{n_p}$ are parameters.
The first term on the right-hand side of \eqref{pg} is equal to the total power absorbed from the WEC dynamics by the PTO at time $t$.  
The second and third terms are used to model transmission losses in the power train between the PTO and the utility bus. 
For an electrical power train, these two terms capture the conductive power dissipation in the PTO and power electronics, with the second term capturing ohmic (i.e., ``$I^2R$'') losses, and the third term capturing diode losses.

\subsection{Energy-preserving discretization}\label{inf_dim_disc}

We assume the control input $\rv{u}(t)$ is implemented as a zero-order hold (ZOH) signal, i.e., 
\begin{equation}\label{ZOH}
\rv{u}(t) = \rv{u}_k , \ \ \forall t \in [kT,(k+1)T)
\end{equation}
where $T$ is the sample time, and $k \in \mathbb{Z}$. 
We analogously refer to continuous-time signals sampled at the transition times of the ZOH mapping via subscripts, i.e., $\rv{d}_k = \rv{d}(kT)$, $\rv{a}_k = \rv{a}(kT)$, and so on.
Let $\rv{p}_k$ be the average value of $\rv{\varrho}(t)$ over the interval $t \in [kT,(k+1)T)$. 
Then 
\begin{equation}
\rv{p}_k = -\left( \rv{u}_k^T \rv{q}_k + \rv{u}_k^T R \rv{u}_k + v_d^T \left| \rv{u}_k \right| \right)
\end{equation}
where 
\begin{equation}\label{qk}
\rv{q}_k \triangleq \tfrac{1}{T} \left( \rv{d}_{k+1} - \rv{d}_k \right).
\end{equation}

The stochastic dynamics of the resulting sampled continuous-time system can be expressed as
\begin{align}
\rv{d}_k =& \left( H_{wd} \rv{w} \right)_k + \left( H_{ud} \rv{u} \right)_k \\
\rv{a}_k =& \left( H_{wa} \rv{w} \right)_k
\end{align}
where $\rv{w}_k \in \mathbb{R}^{n_p+1}$ is an independent, identically-distributed Gaussian stochastic sequence with zero-mean and covariance $S_w \geqslant 0$.  
Mappings $H_{ud}$, $H_{wd}$, and $H_{wa}$ are uniquely characterized by their discrete-time frequency response functions $\bar{H}_{ud}(e^{j\Omega})$, $\bar{H}_{wd}(e^{j\Omega})$ and $\bar{H}_{wa}(e^{j\Omega})$.
It follows that 
\begin{align}
\rv{q}_k =& \left( H_{wq} \rv{w} \right)_k + \left( H_{uq} \rv{u} \right)_k
\end{align}
where mappings $H_{wq}$ and $H_{uq}$ are uniquely characterized by their discrete-time frequency response functions
\begin{align}
\bar{H}_{uq}(e^{j\Omega}) =& \tfrac{1}{T} (e^{j\Omega}-1) \bar{H}_{ud}(e^{j\Omega}) \label{Hbar_ud} \\
\bar{H}_{wq}(e^{j\Omega}) =& \tfrac{1}{T} (e^{j\Omega}-1) \bar{H}_{wd}(e^{j\Omega}) \label{Hbar_wd}
\end{align}

In \cite{lao2022} it was shown that $\bar{H}_{uq}(e^{j\Omega})$, $\Omega \in [-\pi,\pi]$ is
\begin{equation} \label{Huq_1}
\bar{H}_{uq}(e^{j\Omega}) = \sum\limits_{\ell = -\infty}^\infty \hat{Z}_{uv}(j\omega_\ell) \sinc^2(\omega_\ell T/2) 
\end{equation}
where $\omega_\ell \triangleq \Omega/T+2\pi\ell/T$.
$\bar{H}_{ud}(e^{j\Omega})$ is then found via \eqref{Hbar_ud}.
Meanwhile, to get mappings $\hat{H}_{wd}(e^{j\Omega})$ and $\bar{H}_{wa}(e^{j\Omega})$, we first find the discrete-time joint PSD
\begin{equation} \label{Sigma_q_a_1}
\Sigma_{y_f}(\Omega) = \sum\limits_{\ell=-\infty}^\infty \frac{1}{T} \sinc^2(\omega_\ell T/2) S_{y_f}(\omega_\ell).
\end{equation}
Then, via the Spectral Factorization Theorem, we find $\bar{H}_{wd}(e^{j\Omega})$ and $\bar{H}_{wa}(e^{j\Omega})$ as causal, minimum-phase, discrete-time frequency-response functions satisfying
\begin{equation} \label{Sigma_q_a_2}
\Sigma_{y_f}(\Omega) = \begin{bmatrix} \bar{H}_{wd}(e^{j\Omega}) \\ \bar{H}_{wa}(e^{j\Omega}) \end{bmatrix}  S_w
\begin{bmatrix} \bar{H}_{wd}(e^{-j\Omega}) \\ \bar{H}_{wa}(e^{-j\Omega}) \end{bmatrix}^T.
\end{equation}
Frequency response function $\bar{H}_{wq}(e^{j\Omega})$ is then found via \eqref{Hbar_wd}.
We note that in \eqref{Sigma_q_a_2}, the factorization is not unique.
Without loss of generality, we adopt the convention of a canonical factorization \cite{sayed2001}, in which $S_w$ is normalized such that
\begin{equation}
\frac{1}{2\pi}
\int_{-\pi}^\pi \begin{bmatrix} \bar{H}_{wd}(e^{j\Omega}) \\ \bar{H}_{wa}(e^{j\Omega}) \end{bmatrix} d\Omega = I.
\end{equation}

Associated $z$-domain transfer functions $\bar{H}_{uq}(z)$, $\bar{H}_{wq}(z)$, $\bar{H}_{ud}(z)$, $\bar{H}_{wd}(z)$ and $\bar{H}_{wa}(z)$ are obtained by inverse-transforming the discrete-time frequency-response functions to get discrete-time impulse response functions, and then $z$-transforming these.
The following theorem groups together some useful properties of these transfer functions.
\begin{th1}\label{discrete_time_properties_theorem}
If Assumption \ref{assumption_1} holds then:
\begin{enumerate}
\item 
\label{properties_1}
$\{ \bar{H}_{uq}, \bar{H}_{ud} \} \in \mathbb{H}_\infty$ and $\{\bar{H}_{wa}, \bar{H}_{wd} \} \subset \mathbb{H}_2 \cap \mathbb{H}_\infty$.
\item 
\label{properties_2}
$\bar{H}_{uq}$ is positive-real in discrete-time, i.e., it is analytic for all $z \in \mathbb{C}_{>1}$ and satisfies
\begin{equation}\label{properties_theorem_pr}
\bar{H}_{uq}(z) + \bar{H}_{uq}^H(z) \geqslant 0, \ \ \forall z \in \mathbb{C}_{\geqslant 1}
\end{equation}
with the inequality holding strictly except for $z=1$.
\item
\label{properties_6}
$\bar{H}_{uq}(1) = 0$, and $\bar{H}_{ud}(1) + \bar{H}_{ud}^H(1) > 0$. 
\end{enumerate}
\end{th1}

\subsection{Finite-dimensional discrete-time model}\label{finite_dim_disc}

We assume that the mappings $H_{ud}$, $H_{wd}$, and $H_{wa}$ have been approximated as finite-dimensional, LTI, discrete-time systems, using any of several system identification techniques.
Let $\rv{y} \triangleq [\rv{d}^T \ \rv{a}]^T$. 
Without loss of generality we assume the resulting state space system for the mapping $\{\rv{u},\rv{w}\} \mapsto \rv{y}$ is in the form of an innovations model \cite{anderson2012optimal}, i.e., 
\begin{equation}\label{discrete_time_state_space_orig}
H : \left\{ \begin{array}{rl}
\rv{x}_{k+1} =& A \rv{x}_k + B_u \rv{u}_k + B_w \rv{w}_k \\
\rv{y}_k  =& C_y \rv{x}_k + \rv{w}_k
\end{array} \right.
\end{equation}
and a minimal realization%
\footnote{Note that, as with all innovations models, the state vector $\rv{x}_k$ should not be thought of as a physical state, but rather, as the optimal state estimate for the system, conditioned on $\{\rv{y}_\ell, \ell < k\}$. Model \eqref{discrete_time_state_space_orig} can equivalently be viewed as a Kalman filter, with matrix $B_w$ as the associated Kalman gain.
}.
For convenience, define partitions
\begin{equation}
C_y = \begin{bmatrix} C_d \\ C_a \end{bmatrix}, \quad
I = \begin{bmatrix} D_{wd} \\ D_{wa} \end{bmatrix}
\end{equation}
and note that
\begin{align}
\label{Hud_approx}
\bar{H}_{ud}(z) \approx & C_{d} \left[ zI - A \right]^{-1} B_u \\
\label{Hwd_approx}
\bar{H}_{wd}(z) \approx & D_{wd} + C_{d} \left[ zI - A \right]^{-1} B_w \\
\label{Hwa_approx}
\bar{H}_{wa}(z) \approx & D_{wa} + C_a \left[ zI - A \right]^{-1} B_w.
\end{align}
Furthermore, $\rv{q}_k$ is 
\begin{equation}
\rv{q}_k = C_q \rv{x}_k + D_{uq} \rv{u}_k + D_{wq0} \rv{w}_k + D_{wq1} \rv{w}_{k+1}
\end{equation}
where
\begin{align}
C_q =& \tfrac{1}{T} C_d (A-I), & D_{uq} =& \tfrac{1}{T} C_d B_u \\
D_{wq0} =& \tfrac{1}{T} (C_dB_w-D_{wd}) , & D_{wq1} =& \tfrac{1}{T} D_{wd}
\end{align}
Consequently, we may express transfer functions $\bar{H}_{uq}$ and $\bar{H}_{wq}$ in terms of the state space parameters as
\begin{align}
\label{Huq_approx}
\bar{H}_{uq}(z) \approx & D_{uq} + C_q [zI-A]^{-1}B_u \\
\label{Hwq_approx}
\bar{H}_{wq}(z) \approx & D_{wq1}z + D_{wq0} + C_q [zI-A]^{-1} B_w.
\end{align}

We presume that the finite-dimensional model has been found in a way that preserves the properties listed in Theorem \ref{discrete_time_properties_theorem} for the above transfer functions.
More information on subspace-based procedures to identify the state space parameters of $H$, while adhering to the conditions of Theorem \ref{discrete_time_properties_theorem}, can be found in \cite{hoagg2004} and \cite{lao2019}.
The following theorem groups together some useful implications of Theorem \ref{discrete_time_properties_theorem}, which apply to the finite-dimensional model $H$ as in \eqref{discrete_time_state_space_orig}.

\begin{th1} \label{finite_dimensional_model_properties_theorem}
Assume the minimal finite-dimensional model \eqref{discrete_time_state_space_orig} is such that the transfer functions \eqref{Hud_approx}, \eqref{Hwd_approx}, \eqref{Hwa_approx}, and \eqref{Huq_approx} adhere to the conditions in Theorem \ref{discrete_time_properties_theorem}. Then:
\begin{enumerate}
\item \label{Schur_property}
$A$ is Schur. 
\item \label{Feedthrough_property}
$D_{uq}+D_{uq}^T > 0$. 
\item \label{KYP_lemma_property}
Let $E\in\mathbb{R}^{n_o\times n}$ be full-row-rank with null space equal to the  unobservable subspace of $(A,C_q)$.
Then the discrete-time Riccati equation
\begin{equation}\label{KYP_riccati}
W = A^T W A + F^T M F
\end{equation}
with
\begin{align}
\label{M}
M  \triangleq & R + \tfrac{1}{2}(D_{uq}+D_{uq}^T) - B_u^TWB_u  \\
\label{F} 
F \triangleq & -M^{-1} \left[ \tfrac{1}{2}C_q-B_u^TWA \right]
\end{align}
has a solution $W=W^T\geqslant 0$ with $EWE^T>0$, $M > 0$, and such that $A+B_uF$ is Schur.
\end{enumerate}
\end{th1}

\subsection{Inverse system dynamics}\label{stall_section}

Consider the inverse system in which $\rv{d}$ is treated as a control input, and $\rv{u}$ is treated as an output. 
Then
\begin{align}
\rv{u}_k =& -D_{uq}^{-1} C_q \rv{x}_k - \tfrac{1}{T} D_{uq}^{-1} \left( C_d B_w - D_{wd} \right) \rv{w}_k
\nonumber \\ 
& - \tfrac{1}{T} D_{uq}^{-1}D_{wd} \rv{w}_{k+1} + \tfrac{1}{T} D_{uq}^{-1} (\rv{d}_{k+1}-\rv{d}_k) .
\label{inverse_system_u_eq1}
\end{align}
As such, 
\begin{multline} \label{inverse_system_xi_deq}
\rv{x}_{k+1} = \left[ A - B_u D_{uq}^{-1} C_q \right] \rv{x}_k + \left[ \tfrac{1}{T} B_u D_{uq}^{-1} \right] \left( \rv{d}_{k+1} -\rv{d}_{k} \right) \\
+ \tfrac{1}{T} B_u D_{uq}^{-1} D_{wd} (\rv{w}_k - \rv{w}_{k+1}) \\
+ [I - \tfrac{1}{T} B_u D_{uq}^{-1} C_d ] B_w \rv{w}_k.
\end{multline} 
Changing coordinates, define $\rv{\zeta}_k$ such that
\begin{equation}
\begin{bmatrix} \rv{d}_k \\ \rv{\zeta}_k  \end{bmatrix} = \begin{bmatrix} C_d \\ B_u^\perp \end{bmatrix} \rv{x}_k + \begin{bmatrix} D_{wd}  \\ 0 \end{bmatrix} \rv{w}_k
\end{equation}
where $B_u^\perp$ is a full-row-rank matrix such that $B_u^\perp B_u = 0$ and such that $[C_d^T \ (B_u^\perp)^T]$ is square and invertible.
(Note that this is assumed without loss of generality, because $C_dB_u = T D_{uq}$ is nonsingular.)
Then we have that
\begin{equation}
\begin{bmatrix} C_d \\ B_u^\perp \end{bmatrix}^{-1}
=
\begin{bmatrix} B_u & C_d^\perp \end{bmatrix} 
\begin{bmatrix} C_d B_u & 0 \\ 0 & B_u^\perp C_d^\perp \end{bmatrix}^{-1}
\end{equation}
where $C_d^\perp$ is any full-column rank matrix such that $C_d C_d^\perp = 0$ and such that $[B_u \ C_d^\perp]$ is square.
Using these facts, it is straight-forward to verify that in the new coordinates,
\begin{align}\label{stall_state_system_1}
\rv{\zeta}_{k+1} =& A_s \rv{\zeta}_k  + B_{ws} \rv{w}_k + B_{ds} \rv{d}_k
\\
\label{stall_state_system_2}
\rv{x}_k =& C_s \rv{\zeta}_k + D_{ws} \rv{w}_k + D_{ds} \rv{d}_k
\end{align}
where
\begin{align}
\label{As}
A_s \triangleq & B_u^\perp A C_s \\
\label{Bs}
B_{ws} \triangleq & B_u^\perp B_w - B_{ds} D_{wd}, &
B_{ds} \triangleq & B_u^\perp A D_{ds} \\
\label{Cs}
C_s \triangleq & C_d^\perp \left[ B_u^\perp C_d^\perp \right]^{-1} \\
\label{Ds}
D_{ws} \triangleq & - D_{ds} D_{wd}, &
D_{ds} \triangleq & \tfrac{1}{T} B_u D_{uq}^{-1} 
\end{align}

\begin{th1} \label{Schur_theorem}
If the conditions of Theorems \ref{discrete_time_properties_theorem} and \ref{finite_dimensional_model_properties_theorem} hold, then  $A_s$ is Schur.
\end{th1}

\section{Optimal Stochastic WEC Control}

Let $\bm{Y}_k$ be
\begin{equation}
\bm{Y}_k \triangleq \left\{ \rv{y}_\ell : \ell < k \right\}
\end{equation}
Then we presume that at time $k$, the information available for the purposes of control is $\bm{Y}_{k}$, i.e., we presume a control algorithm $K$ that facilitates the mapping
\begin{equation}\label{Kc}
K: \bm{Y}_{k} \mapsto \rv{u}_k
\end{equation}
for each $k \in \mathbb{Z}_{\geqslant 0}$, starting from a deterministic initial condition $\rv{x}_0 = x_0$.
Random variables $\rv{x}_k$ and $\rv{u}_k$ are functions of $\bm{Y}_k$.
At time $k$ the data $\rv{Y}_k = Y_k$ is known to the control algorithm $K$, and consequently the data $\rv{x}_k = x_k$ and $\rv{u}_k = u_k$ are known.
With each time advancement of $k \rightarrow k+1$, innovations data $\rv{w}_k = w_k$ becomes known, and data $\rv{x}_{k+1} = x_{k+1}$ is evaluated recursively by control algorithm $K$, via \eqref{discrete_time_state_space_orig}. 

Let $\mathbb{K}_c$ be the set of all strictly causal mappings $K$ as in \eqref{Kc}. 
The idealized optimal WEC control problem may be stated as the following optimization problem:
\begin{equation} \label{Kopt_1}
K = \sol \left\{ \begin{array}{rl}
\text{Given}: & A, B_u, B_w, C_y, S_w, x_0 \\
\text{Max}: & p_{\text{avg}} \\
\text{Dom}: & K \in \mathbb{K}_c \\
\text{Constr}: & |\rv{u}_{k}| \leqslant u_{\max} \wpone,  \ \forall k\in\mathbb{Z}_{\geqslant 0}  \\
&  |\rv{d}_{k}| \leqslant d_{\max} \wpone, \ \forall k\in\mathbb{Z}_{>0}
\end{array} \right.
\end{equation}
where $u_{\max}$ and $d_{\max}$ are the vectors of force and displacement limits for PTOs, and 
\begin{align}
p_{\text{avg}} 
\triangleq & \lim\limits_{\tau\rightarrow\infty} \frac{1}{\tau+1} \sum\limits_{k=0}^\tau \mathcal{E}\{\rv{p}_k\} \\
=& \lim\limits_{\tau\rightarrow\infty} \frac{-1}{\tau+1} \sum\limits_{k=0}^\tau \mathcal{E} \left\{ \rv{u}_k^T C_q \rv{x}_k + \rv{u}_k^T \tilde{R} \rv{u}_k + v_d^T |\rv{u}_k| \right\}
\end{align}
where $\tilde{R} \triangleq R + \tfrac{1}{2}(D_{uq}+D_{uq}^T) > 0$. 
However, we note that optimization problem \eqref{Kopt_1} is not well-posed.
This is because for each time $k$, there is a nonzero probability that there will exist no control input that can simultaneously satisfy both the displacement and force constraints. 
To rectify this, we soften the force constraint using a vector of slack variables, denoted $\rv{b} \in \mathbb{R}^{n_p}_{\geqslant 0}$.  We then have the relaxed problem
\begin{equation}
K = \sol \left\{ \begin{array}{rl}
\text{Given}: & A, B_u, B_w, C_y, S_w, x(0) \\
\text{Max}: & p_{\text{avg}} - \lim\limits_{\tau\rightarrow\infty} \tfrac{1}{\tau+1} \sum_{k=0}^\tau \mathcal{E} \left\{ \mu u_{\max}^T \rv{b}_k \right\} \\
\text{Dom}: & K \in \mathbb{K}_c \\
\text{Constr}: & |\rv{u}_{k}| \leqslant u_{\max} + \rv{b}_{k} \wpone, \ \forall k\in\mathbb{Z}_{\geqslant 0} \\
& |\rv{d}_{k}| \leqslant d_{\max} \wpone, \ \forall k\in\mathbb{Z}_{>0}  \\
& \rv{b}_k \geqslant 0 \wpone, \ \forall k \in \mathbb{Z}_{\geqslant 0}
\end{array} \right.
\end{equation}
where $\mu \in \mathbb{R}_{>0}$ is a penalty term. 

The following theorem is instrumental to the formulation of this control problem in the context of MPC, because it illustrates that the maximization of $p_{\text{avg}}$ may be framed equivalently as the minimization of an auxiliary positive-semidefinite objective function.

\begin{th1}\label{optimal_control_theorem}
Let $K \in \mathbb{K}_c$ be such that the limit
\begin{equation}\label{optimal_control_theorem_eq1}
\lim\limits_{k\rightarrow\infty} 
\frac{1}{k} \mathcal{E} \left\{ \| \rv{x}_k \|^2 \right\} = 0
\end{equation}
Then
\begin{align}
p_{\text{avg}} = & \bar{p} - \lim\limits_{\tau\rightarrow\infty} \frac{1}{\tau+1} \sum\limits_{k=0}^\tau \mathcal{E} \left\{ \| \rv{\xi}_k \|_M^2  + v_d^T |\rv{u}_k| \right\} 
\end{align}
where $\rv{\xi}_k \triangleq \rv{u}_k-F\rv{x}_k$, matrices $M$ is as $F$ are defined in \eqref{M} and \eqref{F}, 
$\bar{p} \triangleq \trace\{ B_w^TWB_wS_w\}$, and $W=W^T\geqslant 0$ is the unique stabilizing solution to \eqref{KYP_riccati}. 
\end{th1}

Using this theorem, we can re-express the control problem for $K$ as the equivalent problem below:
\begin{equation} \label{OCP}
K = \sol \left\{ \begin{array}{rl}
\text{Given}: & A, B_u, B_w, C_y, S_w, x(0) \\
\text{Min}: & \lim\limits_{\tau\rightarrow\infty} \frac{1}{\tau+1} \sum\limits_{k=0}^\tau \mathcal{E} \left\{ J(\rv{x}_k,\rv{u}_k,\rv{b}_k) \right\} \\
\text{Dom}: & K \in \mathbb{K}_c \\
\text{Constr}: & |\rv{u}_{k}| \leqslant u_{\max} + \rv{b}_{k} \wpone,  \forall k\in\mathbb{Z}_{\geqslant 0} \\
& |\rv{d}_{k}| \leqslant d_{\max} \wpone,  \forall k\in\mathbb{Z}_{>0} \\
& \rv{b}_k \geqslant 0 \wpone, \ \forall k \in \mathbb{Z}_{\geqslant 0}
\end{array} \right.
\end{equation}
where $J(x_k,u_k,b_k)$ is equal to 
\begin{equation}
J(x_k,u_k,b_k)
\triangleq \| u_k-Fx_k \|_M^2 
+ v_d^T |u_k| + \mu u_{\max}^T b_k 
\label{J}
\end{equation}
Note that because $M > 0$ and $\mu > 0$, $J$ is a positive-semidefinite, and is convex.

\begin{rem1}
Technically, the optimal control problem \eqref{OCP} is not well-posed. Satisfaction of constraint $|\rv{d}_k| \leqslant d_{\max}$ must be enforced at previous time step, $k-1$, by appropriate choice of $\rv{u}_{k-1}$.
Ensuring satisfaction of the constraint would require knowledge of innovations vectors $\rv{w}_k$ and $\rv{w}_{k-1}$ and state vector $\rv{x}_{k-1}$.
Due to the fact that $K$ is constrained to be strictly causal, this knowledge is not available to the controller until after input $\rv{u}_{k-1}$ has been applied.  
As such, there is a finite (albeit very small, in practice) probability of displacement violation, which is unavoidable. 
To make the optimization \eqref{OCP} well-posed, the displacement constraint can be replaced with a conditional expectation, as
\begin{equation}
\left| \mathcal{E} \left\{ \rv{d}_{k+1} \ | \ \bm{Y}_{k} \right\} \right| \leqslant d_{\max} \wpone, \forall k \in \mathbb{Z}_{\geqslant 0}
\end{equation}
This is what is done in the proposed MPC algorithm.
\end{rem1}

\section{MPC Formulation}

To frame the optimal stochastic WEC control problem in the context of MPC, suppose the present time is $k$, let $h > 0$ be some receding horizon length, and presume that the data $\rv{Y}_k = Y_k$ is known.
Assume that for all time $\ell < k$, the mapping $K : \rv{Y}_\ell \mapsto \rv{u}_\ell$ is imposed, where $K \in \mathbb{K}_c$. 
It follows that at time $k$, the data $\rv{x}_k = x_k$ is also known.

Let the future control inputs over the receding horizon be
\begin{equation}
\rv{U}_k \triangleq \{ \rv{u}_k,...,\rv{u}_{k+h} \}
\end{equation}
and let
\begin{equation}
U_{k|k} \triangleq \{u_{k|k},..., u_{k+h|k}\}
\end{equation}
be a sequence of hypothetical, deterministic future control inputs, which are determined from data $Y_k$ and therefore are probabilistically independent of innovations $\{\rv{w}_k, ..., \rv{w}_{k+h}\}$. 
Let $\{x_{k|k},...,x_{k+h|k}\}$ be the expected values of $\{\rv{x}_k,...,\rv{x}_{k+h}\}$ conditioned on data $Y_k$ and $U_{k|k}$, i.e., 
\begin{equation}
x_{m|k} \triangleq \mathcal{E} \left\{  \rv{x}_m \Big| \rv{Y}_k = Y_k, \rv{U}_k = U_{k|k} \right\} 
\end{equation}
Then it follows that for $m \in \{k,...,k+h\}$, 
\begin{equation}\label{xi_m|k_dynamics}
x_{m+1|k} = A x_{m|k} + B_u u_{m|k}
\end{equation} 
with initial condition $x_{k|k} = x_k$.

In the proposed stochastic MPC formulation of the optimal control problem, at each time $k$ we find a deterministic input sequence $U_{k|k}$ and a deterministic slack variable sequence 
\begin{equation}
B_{k|k} \triangleq \{b_{k|k},...,b_{k+h,k}\}
\end{equation}
which minimize an objective $\Gamma(x_k,U_{k|k},B_{k|k})$, subject to constraints.
As such, we have that
\begin{equation}\label{MPC_algorithm_1}
u_{k}
= \sol\limits_{u_{k|k}} \left\{ \begin{array}{rl}
\text{Given:} & x_k  \\
\text{Min:} & \Gamma(x_k,U_{k|k},B_{k|k}) \\
\text{Dom:} & U_{k|k},B_{k|k} \\
\text{Constr:} & \Theta(x_k,U_{k|k},B_{k|k}) \leqslant 0 \\
& \Psi(x_k,U_{k|k}) = 0
\end{array} \right.
\end{equation}
for appropriately-defined constraint vector functions $\Theta(\cdot,\cdot,\cdot)$ and $\Psi(\cdot,\cdot)$. 
Let the resultant optimized trajectories be denoted by ring accents, i.e., $\mathring{U}_{k|k} = \{ \mathring{u}_{k|k}, ... , \mathring{u}_{k+h|k} \}$ and $\mathring{B}_{k|k} = \{ \mathring{b}_{k|k}, ..., \mathring{b}_{k+h|k} \}$. 
Upon solving this optimization, the MPC algorithm implements $u_k = \mathring{u}_{k|k}$. 
Following this, data $\rv{y}_k = y_k$ is obtained, and innovations data $\rv{w}_k = w_k$ and state data $\rv{x}_{k+1} = x_{k+1}$ are found from \eqref{discrete_time_state_space_orig}. 
Time is incremented, i.e., $k \rightarrow k+1$, and the process is repeated ad infinitum. 

\subsection{Optimization objective}

The MPC optimization objective involves the expected value of $J(\rv{x}_m,u_{m|k},b_{m|k})$ for $m \in \{k,...,k+h\}$, conditioned on data $\rv{Y}_k = Y_k$ as well as $\rv{U}_{k} = U_{k|k}$.
Because $J(\cdot,\cdot,\cdot)$ is quadratic its first argument, the second and third arguments are deterministic when conditioned on the data, and because $x_{m|k}$ is an unbiased conditional estimate of $\rv{x}_m$, it follows that 
\begin{multline}
\label{MPC_objective_eq1}
\mathcal{E} \left\{ \left. J(\rv{x}_m,u_{m|k},b_{m|k}) \right| \rv{Y}_k=Y_k , \rv{U}_k = U_{k|k} \right\} \\
= 
\mathcal{E} \left\{ \| F (\rv{x}_m-x_{m|k})  \|_M^2 \right\} 
+ J(x_{m|k},u_{m|k},b_{m|k}) 
\end{multline}
The MPC optimization objective is the sum of this expectation over the time horizon, plus a final-value penalty.
Noting that $\rv{x}_m-x_{m|k}$ is zero for $m = k$ and a linear combination of $\{\rv{w}_k,...,\rv{w}_{m-1}\}$ for $m>k$, it follows that the first term on the right-hand side of \eqref{MPC_objective_eq1} is independent of optimization variables $U_{k|k}$ and $B_{k|k}$, and can be subtracted from the objective without affecting the solution.
In so doing, the objective to be minimized in the MPC algorithm is 
\begin{multline}
\Gamma(x_k, U_{k|k}, B_{k|k} )  \\
\triangleq J_f(x_{k+h+1|k}) + \sum\limits_{m=k}^{k+h} J(x_{m|k},u_{m|k},b_{m|k}) 
\label{Gamma_def}
\end{multline}
where for $m \in \{k,...,k+h\}$, \eqref{xi_m|k_dynamics} is tacitly assumed, with initial condition $x_{k|k} = x_k$.
Term $J_f(\cdot)$ is the final-value penalty, which will be formulated in Section \ref{sec:final_value}.

\subsection{Constraints}

Inequality constraint vector $\Theta(x_k,U_{k|k},B_{k|k})$ is formulated to impose the constraints on force, displacement, and the slack variables. Specifically, we have that
\begin{equation}
\Theta(x_k,U_{k|k},B_{k|k})
= \begin{bmatrix} 
| u_{k|k} |  - b_{k|k} - u_{\max}
\\ \vdots \\
|u_{k+h|k}| - b_{k+h|k} - u_{\max} 
\\
|d_{k+1|k} |- d_{\max} 
\\ \vdots \\
|d_{k+h|k}| - d_{\max} 
\\
-b_{k|k} 
\\ \vdots \\
-b_{k+h|k}
\end{bmatrix} 
\end{equation}
where
\begin{align}
d_{m|k} \triangleq & \mathcal{E} \left\{ \rv{d}_m | \rv{Y}_k = Y_k, \rv{U}_k = U_{k|k} \right\}
= C_d x_{m|k}
\label{dm|k}
\end{align}
with \eqref{xi_m|k_dynamics} tacitly assumed with initial condition $x_{k|k} = x_k$.

Equality constraint vector $\Psi(x_k,U_{k|k})$ plays an important role in guaranteeing closed-loop stability of the MPC algorithm.
Specifically we impose the condition 
\begin{equation} \label{equality_constraint}
d_{k+h+1|k} = 0.
\end{equation}
As such, $\Psi(x_k,U_{k|k}) = C_d x_{k+h+1|k}$,
with \eqref{xi_m|k_dynamics} tacitly assumed with initial condition $x_{k|k} = x_k$.

\subsection{Final value penalty} \label{sec:final_value}

Due to the imposition of equality constraint \eqref{equality_constraint}, it is known for all $U_{k|k}$ satisfying the constraints of MPC algorithm \eqref{MPC_algorithm_1}, $d_{k+h+1|k} = 0$.
Now, let hypothetical inputs $u_{m|k}$ beyond the time horizon, i.e., for $m > k+h$, be those that maintain $d_{m+1|k} = 0$. 
Using Section \ref{stall_section}, this implies post-horizon trajectories for $x_{m|k}$ and $u_{m|k}$ as
\begin{align}
x_{m|k} =& C_s A_s^{m-(k+h+1)} B_u^\perp x_{k+h+1|k}
\label{stall_state_transition} \\
u_{m|k} =& -D_{uq}^{-1} C_q x_{m|k}.
\label{stall_control_input}
\end{align}
An associated feasible post-horizon penalty term $b_{m|k}$ can then found for all $m \geqslant k+h+1$ as the minimum value that satisfies the force constraint, i.e., 
\begin{equation}
b_{m|k} = \textrm{max} \{ 0,  | u_{m|k} | - u_{\max} \}.
\label{stall_slack_variable}
\end{equation}

Now, for $u_{m|k}$, and $b_{m|k}$ equal to \eqref{stall_control_input} and \eqref{stall_slack_variable} respectively, consider the post-horizon performance measure
\begin{equation}
\label{rho}
\rho(x_{k+h+1|k}) \triangleq \sum\limits_{m=k+h+1}^\infty J(x_{m|k},u_{m|k},b_{m|k})
\end{equation}
Ideally, we would use $\rho(\cdot)$ as the final-value penalty $J_f(\cdot)$ in MPC performance measure \eqref{Gamma_def}.
However, $\rho(x_{k+h+1|k})$ is a complex function of its argument, and in the lemma below, we show that it may be over-bounded by a much simpler function.

\begin{lm1}\label{Jf_lemma}
There exists a final-value penalty function $J_f(\cdot)$ of the form
\begin{equation}
\label{Jf_lemma_def}
J_f(x_{k+h+1|k}) = \| x_{k+h+1|k} \|_{Q_f}^2 + \| C_f x_{k+h+1|k} \|_1
\end{equation}
where $C_f \in \mathbb{C}^{n_p\times n}$ and $Q_f \in \mathbb{R}^{n\times n}$ with $Q_f = Q_f^T \geqslant 0$, such that the for all $x_{k+h+1|k} \in \mathbb{R}^n$, the following hold:
\begin{align}
\label{Jf_lemma_ineq1}
& \rho(x_{k+h+1|k}) \leqslant J_f(x_{k+h+1|k}) \\
\label{Jf_lemma_ineq2}
& J_f(x_{k+h+2|k}) + J(x_{k+h+1|k},u_{k+h+1|k},b_{k+h+1|k}) \nonumber \\ 
& \quad\quad\quad\quad\quad \leqslant J_f(x_{k+h+1|k})
\end{align}
where for the second inequality, $x_{m|k}$, $u_{m|k}$, and $b_{m|k}$ are evaluated as in \eqref{stall_state_transition}, \eqref{stall_control_input}, and \eqref{stall_slack_variable} respectively.
\end{lm1}

We note that the construction of $C_f$ and $Q_f$ is given explicitly in the proof to Lemma \ref{Jf_lemma}, in the appendix.

\subsection{Stability}

In this subsection we provide a stability proof for MPC algorithm \eqref{MPC_algorithm_1}.
To be more concise, we introduce the notation
\begin{align}
\| x \|_{1\tau} \triangleq & \sum\limits_{k=0}^\tau \| x_k \|_1, & 
\| x \|_{2\tau} \triangleq & \left( \sum\limits_{k=0}^\tau \| x_k \|_2^2 \right)^{1/2}
\end{align}
with similar notation for norms on subsequences of $w$.  
Using this notation, the theorem below distills to several common notions of stability as special cases.  
For the case in which $w_k=0$ for all $k\geqslant 0$ and initial condition $x_0 \neq 0$, the theorem guarantees that $\|x\|_{2\tau}$ is bounded from above as $\tau \rightarrow \infty$, thus guaranteeing that $\|x_k\|_2 \rightarrow 0$ as $k \rightarrow \infty$.
This implies asymptotic Lyapunov stability. 
For the case in which $w_k\neq 0$ for some $k\geqslant 0$, and $x_0 = 0$, the theorem establishes a bound on $\| x \|_{2\tau}$ which is a continuous function of $\| w \|_{1\tau}$ and $\| w \|_{2\tau}$, which may be interpreted as a form of bounded-input bounded-state stability. 
Furthermore, if $\|\tfrac{1}{1+\tau} w\|_{1\tau}$ and $\| \tfrac{1}{1+\tau} w\|_{2\tau}$ have finite upper bounds as $\tau \rightarrow \infty$, then the result of the theorem guarantees a bound on  $\lim_{\tau\rightarrow\infty} \| \tfrac{1}{1+\tau} x \|_{2\tau}$.
This has a relationship to mean-square stability for the stochastic system model, as will be shown in the corollary after the theorem. 

\begin{th1}\label{stability_theorem}
Let $\tau \in \mathbb{Z}_{\geqslant 0}$.
Assume that for all $k \in \{0,...\tau\}$, $K : Y_k \mapsto u_k$ is the MPC algorithm \eqref{MPC_algorithm_1}. 
Let $w$ be any input sequence with $\|w_k\|$ bounded for all $k\in\{0,..,\tau\}$ and let $x_0$ be any bounded initial condition.
Then there exist constants $\{\alpha_1...,\alpha_4\} \subset \mathbb{R}_{>0}$ such that
\begin{equation} \label{stability_theorem_bound}
\| x \|_{2\tau}^2 \leqslant \alpha_1 \|x_0\|_1 \\ + \alpha_2 \|x_0\|_2^2 + \alpha_3 \|w\|_{1\tau} + \alpha_4 \|w\|_{2\tau}^2
\end{equation}
\end{th1}

For the stochastic model we consider in this paper, the above theorem may be interpreted as follows. 
Consider that property (c) above holds for all stochastic sequences $\rv{w} = w$ and all initial conditions $\rv{x}_0 = x_0$ in the ensemble. Consequently, we have that
\begin{equation}
\tfrac{1}{1+\tau} \| \rv{x} \|_{2\tau}^2 \leqslant 
\tfrac{\alpha_1}{1+\tau} \| \rv{x}_0 \|_1 + \tfrac{\alpha_2}{1+\tau} \| \rv{x}_0 \|_2^2 + \tfrac{\alpha_3}{1+\tau} \| \rv{w} \|_{1\tau} + \tfrac{\alpha_4}{1+\tau} \| \rv{w} \|_{2\tau}^2 
\end{equation}
Taking $\tau \rightarrow \infty$, we have that 
\begin{align}
&\lim \limits_{\tau\rightarrow\infty} \tfrac{1}{1+\tau} \|\rv{x} \|_{2\tau}^2 
\leqslant \lim\limits_{\tau\rightarrow\infty} \tfrac{1}{1+\tau} \left( \alpha_3 \| \rv{w} \|_{1\tau} + \alpha_4 \| \rv{w} \|_{2\tau}^2 \right) 
\end{align}
But recall that each $\rv{w}_k$ is an independent, identically-distributed, Gaussian random variable with zero mean and covariance $S_w$.  Via the Strong Law of Large Numbers, we therefore have that the following limits hold:
\begin{align}
\label{limit_1}
\lim\limits_{\tau\rightarrow\infty} \tfrac{\alpha_3}{1+\tau} \| \rv{w} \|_{1\tau} =& \alpha_3 \sum\limits_{i=1}^{n_p} \sqrt{\tfrac{2}{\pi} \hat{e}_i^T S_w \hat{e}_i },  \ \ \as \\
\label{limit_2}
\lim\limits_{\tau\rightarrow\infty} \tfrac{\alpha_4}{1+\tau} \| \rv{w} \|_{2\tau}^2 =& \alpha_4 \trace\{ S_w \} , \ \ \as
\end{align}
We conclude that over the ensemble, there exists a bound $N$ equal to the sum of \eqref{limit_1} and \eqref{limit_2}, such that
\begin{align}
&\lim \limits_{\tau\rightarrow\infty} \tfrac{1}{1+\tau} \|\rv{x} \|_{2\tau}^2 
\leqslant N , \ \ \as
\end{align}
As such, we have that in a stochastic context, MPC feedback law \eqref{MPC_algorithm_1} is mean-square stable, almost-surely.

\section{Example}

\begin{figure}
\centering
\includegraphics[scale=.8]{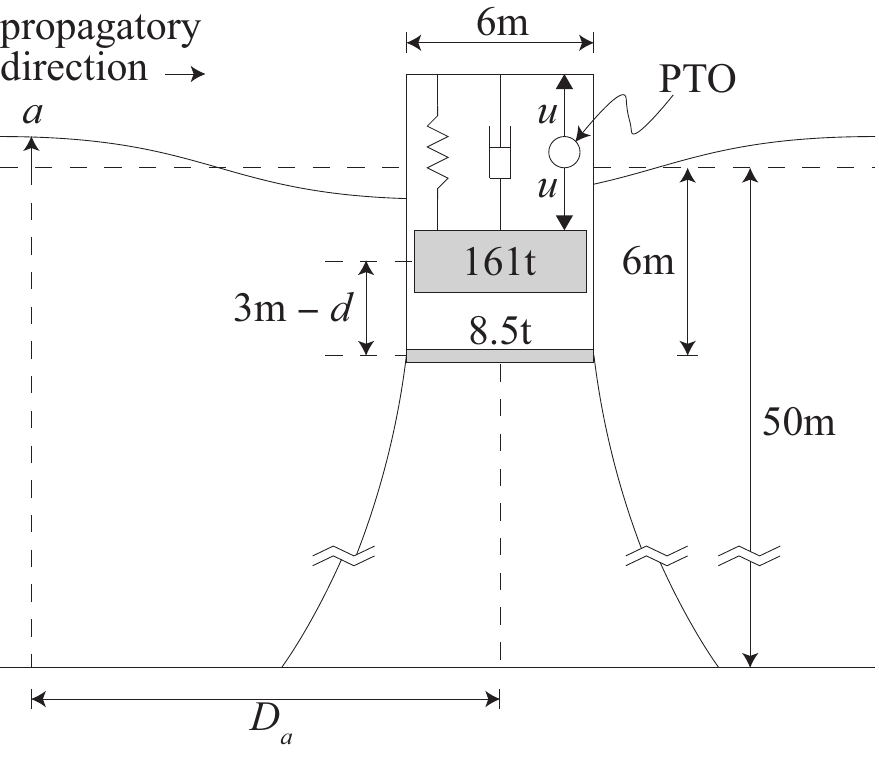}
\caption{Example WEC system}
\label{wec_diagram}
\end{figure}

To demonstrate the application of MPC algorithm \eqref{MPC_algorithm_1}, consider the simple WEC system illustrated in Fig.~\ref{wec_diagram}. 
As shown, the WEC is comprised of a floating, slack-moored, cylindrical buoy, in which a tuned vibration absorber (TVA) is embedded.
The TVA is comprised of a mass-spring-dashpot assembly, and the PTO is situated between the mass and the buoy.
The dynamics of the mass along the buoy axis, together with the heave motion of the buoy, comprise a two degree-of-freedom vibratory system.
We presume that the mooring system restrains the buoy motion to heave, and that the resultant dynamics can justifiably be approximated as linear.

The masses of the buoy and TVA were chosen such that the system is in hydrostatic equilibrium in the configuration shown in Fig.~\ref{wec_diagram}. 
The spring of the TVA is such that the fundamental vibratory mode of the system (including the added mass of the displaced fluid) has a natural period of approximately $9$s.
The dashpot of the TVA is such that the fundamental mode has a fraction of critical damping of $0.5\%$.  

Regarding the PTO model, the scope of this example requires only that we specify its parasitic loss parameters, which were chosen as $R = 1 \text{kW/MN}^2$, and $v_d = 1 \text{kW/MN}$.
(These values were deemed reasonable if the PTO is realized as a permanent-magnet synchronous machine.) 
We presume a force rating of $u_{\max} = 0.5\text{MN}$ and a stroke limit of $d_{\max} = 3\text{m}$. 

The free surface elevation $a(t)$ is presumed to be measured and available for feedback. 
As shown in Fig.~\ref{wec_diagram}, it is presumed that these measurements are made at a distance $D_a$ up-wave from the buoy, along the direction of wave propagation.
The specific value of $D_a$ will be varied in the example, with $D_a = 0$ implying that measurements are colocated with the buoy.
As $D_a$ is increased from $0$, the measurements of $a(t)$ can be thought of as providing a kind of preview of the wave loading the buoy will experience in the near-future. 
However, this is only an approximate interpretation, for two reasons. 
Firstly, the transfer function from the free surface elevation at the the location of the WEC, to the corresponding wave loading force $f(t)$, is noncausal. 
Consequently, even if some preview information were available for this free-surface elevation, it would not be possible to exactly determine a preview of $f(t)$.
Secondly, ocean waves are dispersive, and consequently waves at different temporal frequencies travel at different velocities. 
As such, the time history for the free surface elevation at the buoy cannot be viewed simply as a time-delayed version of $a(t)$ at location $D_a > 0$. 
Despite these two caveats, for the limiting case of $D_a \rightarrow \infty$, knowledge of $a(\tau)$ for all $\tau<t$ implies knowledge of the future incident wave loading $f(\tau)$ over an arbitrary but finite receding horizon $\tau \in [t,t+t_h]$.  

The stochastic sea state was taken to be a Pierson-Moskowitz spectrum \cite{faltinsen1993sea}, i.e.,
\begin{equation}
S_a(\omega) = c_a \left| \tfrac{\omega_p}{\omega} \right|^5 \exp\left\{ -\tfrac{5}{4} \left( \tfrac{\omega_p}{\omega} \right)^4 \right\}
\end{equation}
This spectrum is traditionally parametrized by its significant wave height $H_s$ and peak wave period $T_p$, with these parameters determining $c_a$ and $\omega_p$ in the equation above. 
We have that $T_p = \tfrac{2\pi}{\omega_p}$ and $H_s = 4\sigma_a$, with $\sigma_a^2 = \tfrac{1}{2\pi} \int_{-\infty}^\infty S_a(\omega) d\omega$. 
In this example, we uniformly presume that $T_p = 9$s.  
We take $H_s = 1$m unless otherwise specified. 

\begin{figure}
\centering
\includegraphics[scale=0.75]{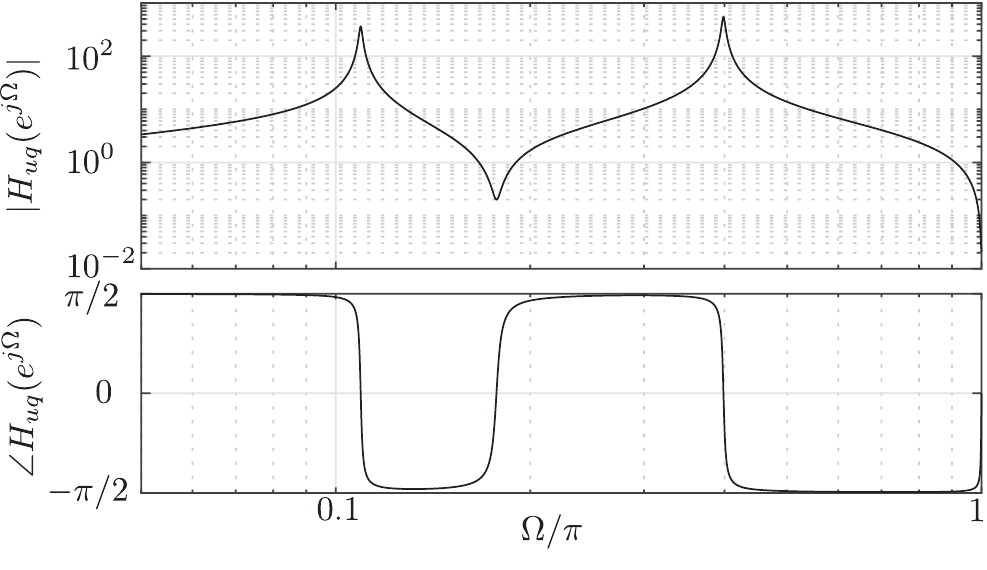}
\caption{Frequency response function $\bar{H}_{uq}(e^{j\Omega})$.  (Both infinite-dimensional model and finite-dimensional approximation are shown, but are indistinguishable.) }
\label{freq_resp}
\end{figure}

\begin{figure}
\centering
\includegraphics[scale=.75]{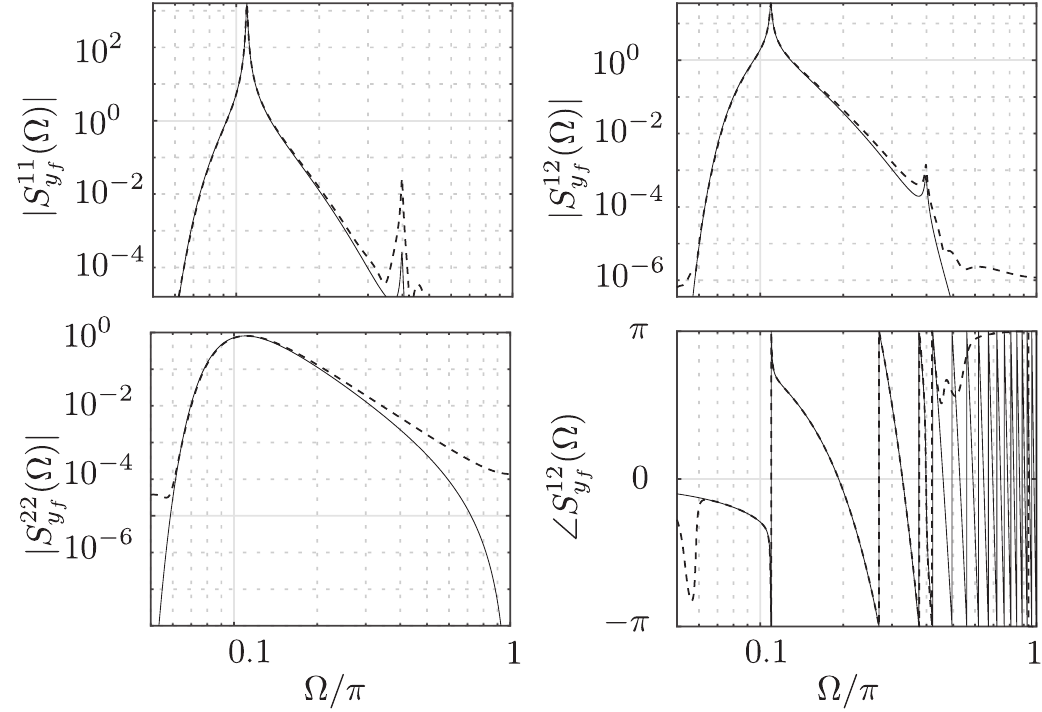}
\caption{Spectrum $\Sigma_{y_f}(\Omega)$ for infinite-dimensional model (solid) and finite-dimensional approximation (dashed). Note that $S_{y_f}^{11}$ and $S_{y_f}^{22}$ have zero phase, and $S_{y_f}^{21}$ and $S_{y_f}^{12}$ are complex conjugates.}
\label{spec_resp}
\end{figure}

The sample time for the controller is uniformly taken to be $T = 0.5\text{s}$.  
Fig.~\ref{freq_resp} shows the discrete-time frequency response function $\bar{H}_{uq}(e^{j\Omega})$, both for the original infinite-dimensional model (as in Section \ref{inf_dim_disc}) and the finite-dimensional approximation (as in Section \ref{finite_dim_disc}). 
Note that the both the infinite-dimensional and finite-dimensional models are positive-real.
Analogously, Fig.~\ref{spec_resp} shows the discrete-time spectrum $\Sigma_{y_f}(\Omega)$, both for the infinite-dimensional model and the finite-dimensional approximation, showing a good match at all frequencies where the spectrum magnitude is significant.
(This spectrum corresponds to $D_a = 20$m.)
It is straight-forward to verify from this plot that the infinite-dimensional model adheres to Assumption \ref{assumption_1}\ref{assumption_1_domains}.


MPC algorithm \eqref{MPC_algorithm_1} was implemented for this model, with a time horizon of one minute (i.e., $h = 120$).  The only other parameter necessary to specify is the slack penalty factor $\mu$.
Performance was found to be insensitive to this value, so long as it is sufficiently large to prohibit violations of the force constraint, and a value of $\mu = 10$ was used for all simulations.
All simulations were performed in Matlab, and the optimizations were implemented using CVX \cite{cvx}.  

\begin{figure*}
\centering
\includegraphics[scale=.8]{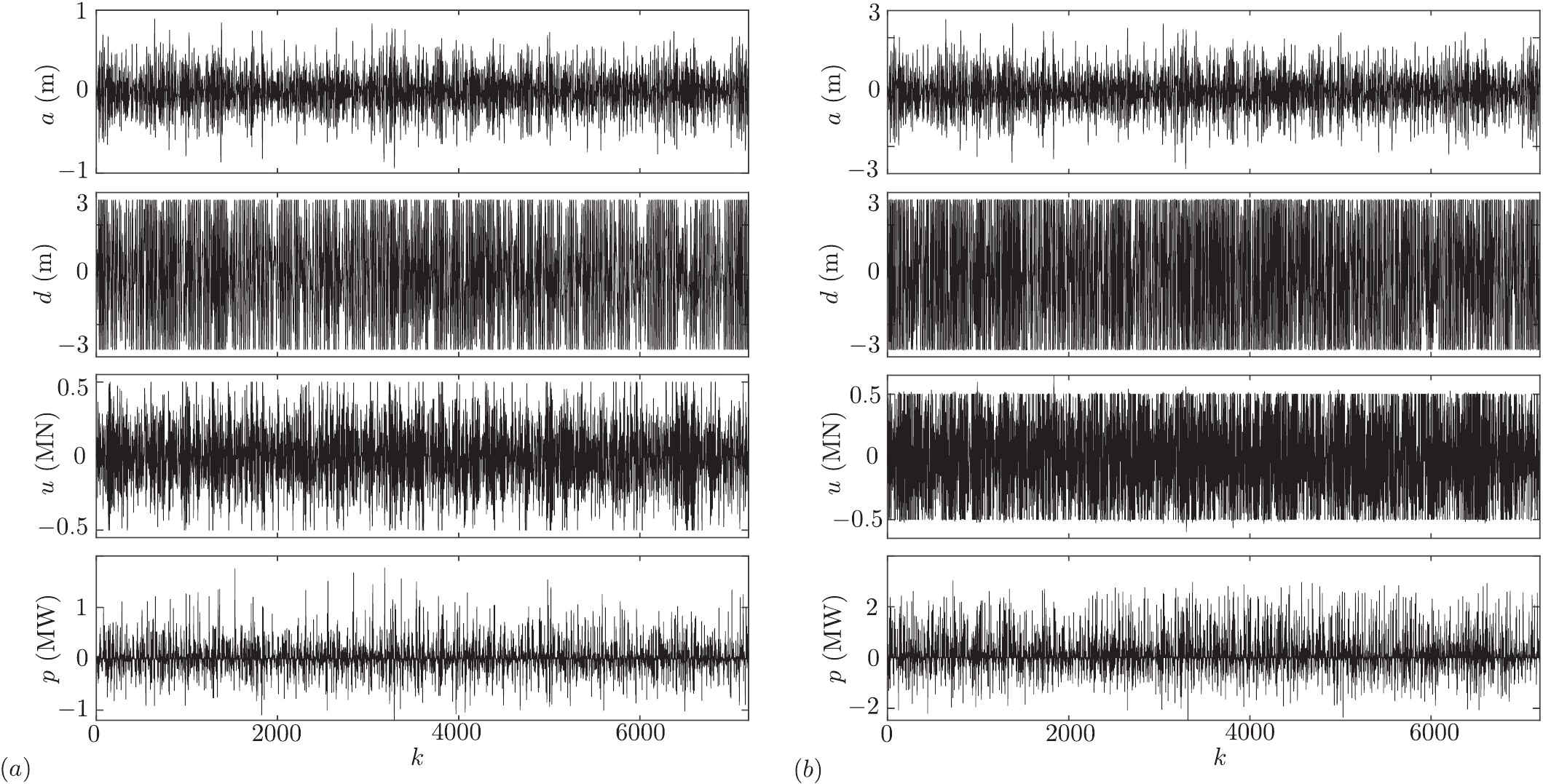}
\caption{Sample path realization for $a_k$, $d_k$, $u_k$, and $p_k$, with MPC algorithm \eqref{MPC_algorithm_1}, for the case with $D_a = 0$m and $H_s = 1$m (a), and $3$m (b)}
\label{transient_plots}
\end{figure*}

Fig.~\ref{transient_plots}a shows the a sample path realization of the stochastic response of the WEC for the case with $D_a = 0$m.  Note that both the force and stroke constraints are uniformly satisfied, as desired.
The generated power $p_k$ is also shown, and it is worth noting that the MPC algorithm requires significant bi-directional power flow (i.e., both significantly positive and significantly negative values). 
Fig.~\ref{transient_plots}b shows analogous plots for the same case but with $H_s = 3$m.  
Note that although the average force magnitude is higher than in the $H_s = 1$m case, the force rating is only violated very occasionally. 
These violations occur when it is infeasible for the MPC algorithm to simultaneously satisfy both the stroke and force constraints.

\begin{figure}
\centering
\includegraphics[scale=.8]{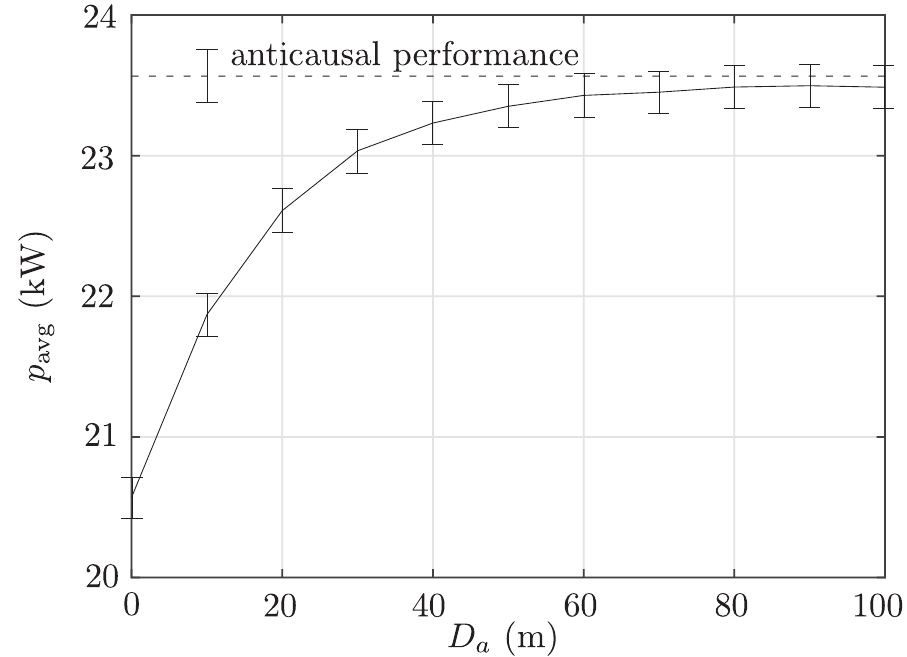}
\caption{Estimates of $p_{\text{avg}}$ (with $90\%$ confidence intervals) vs. $D_a$, with $H_s = 1$m.}
\label{avg_power_plot}
\end{figure}

To estimate the performance $p_{\text{avg}}$ achieved by controller, an ensemble of $48$ sample path realizations was simulated for one hour each.
The average power generation was evaluated for each simulation, resulting in a sample set of $48$ values.
From this sample set, the mean was estimated, and confidence intervals were assessed for this estimate.
The resultant estimates are shown in Fig.~\ref{avg_power_plot} for $H_s = 1$m, and for various values of $D_a$. 
As expected, as $D_a$ is increased, the generated power also increases. 
As $D_a$ approaches $100$m, the performance asymptotically approaches a limit.

Also shown on the plot is the anticausal power generation performance.
This was obtained by optimizing $p_{\text{avg}}$ for each sample in the sample set, assuming the entire $w$ trajectory is known \emph{a priori}. Additionally, in this case the optimal control trajectory was optimized all at once, rather than in the form of an MPC algorithm.
As expected, the anticausal performance exceeds the causal performance for all values of $D_a$.
Theoretically, the anticausal optimal performance should be slightly higher than the asymptote of the causal case, as $D_a$ is made large.
This asymptotic discrepancy is due to a small degree of sub-optimality inherent to the MPC algorithm as a consequence of its finite horizon.
However, it is worth noting that as $D_a$ becomes large, the anticausal and causal performance estimates are close enough that they are within the  margin of error for the simulation.

\begin{figure}
\centering
\includegraphics[scale=.8]{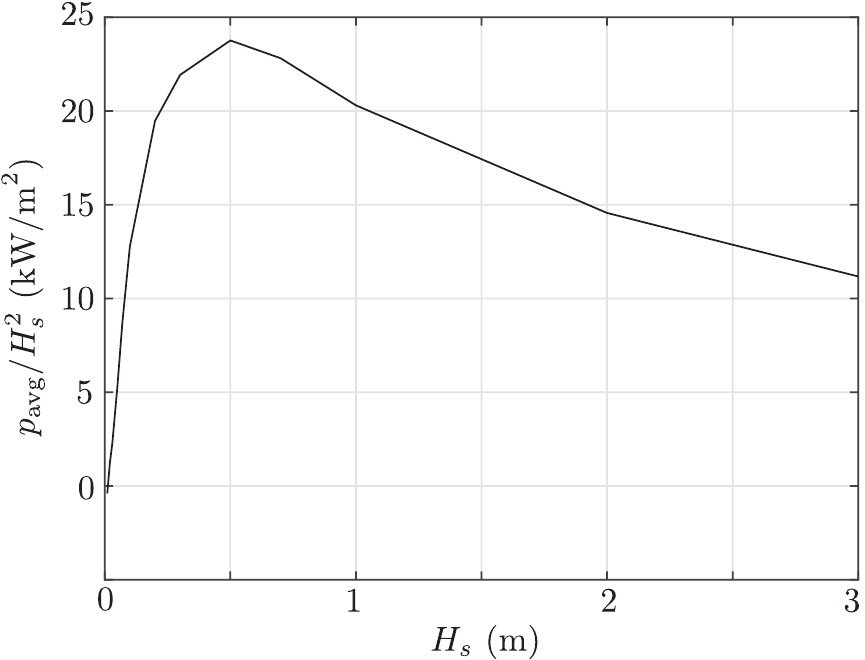}
\caption{Estimates of $p_{\text{avg}}$ as a function of $H_s$, with $D_a = 0$m.}
\label{avg_power_plot_2}
\end{figure}

Finally, we examine the manner in which performance changes with $H_s$.
Fig.~\ref{avg_power_plot_2} shows the estimate for $p_{\text{avg}}$ for $D_a = 0$m, as a function of $H_s$.  To illustrate the features of this plot better, $p_{\text{avg}}$ is normalized by $H_s^2$.  
(Note that for an unconstrained WEC with linear dynamics and linear control, this quantity would be independent of $H_s$.) 
Clearly, we see that, when normalized as such, the power generation performance has a peak at approximately $H_s = 0.5$m.
Above this wave height, the constraints on stroke and force begin to hamper the ability of the WEC to harvest the available energy.
Below this wave height, the efficiency of the WEC power train drops significantly due to the $v_d^T|u|$ term in the expression for the parasitic dissipation.

\section{Conclusions}

In this paper we have illustrated a systematic technique for synthesizing an MPC algorithm for a stochastically-excited WEC, which adheres to force and stroke constraints whenever possible, is provably stable, and does not require the incident wave force to be forecast explicitly.
An interesting extension to this work would be the development of a version of the technique which can accommodate nonlinearities in the WEC dynamics. 
Other items for further study include the stability and performance robustness of the algorithm, in the presence of uncertainty in the WEC dynamic model.

\appendix

\subsection{Proof of Theorem \ref{discrete_time_properties_theorem}}

To begin, we observe that $\bar{H}_{uq}(z)$ is analytic on $\mathbb{C}_{>1}$.
Let $u^i_k = \hat{e}_i \delta_{k}$, where $\hat{e}_i$ is the Cartesian unit vector in the $i^{th}$ direction and $\delta_{k}$ is the Kronecker delta, and let $u^i(t)$ be the resultant continuous-time ZOH input, via \eqref{ZOH}. Let $v^i(t)$ be the resultant continuous-time response, i.e., $v^i(t) = (Z_{uv}u^i)(t)$.
Let $q^i_k$ be the corresponding discrete-time response.
Then we note that 
$
\sup_{k\in\mathbb{Z}} \| q^i_k \|_2 \leqslant \| v^i \|_{\mathbb{L}_\infty}.
$
It is a standard result from input-output analysis that
\begin{align*}
\| v^i \|_{\mathbb{L}_\infty} & \leqslant \| \tilde{Z}_{uv} \|_{\mathbb{L}_1} \| u^i \|_{\mathbb{L}_\infty} = \| \tilde{Z}_{uv} \|_{\mathbb{L}_1} \times 1
\end{align*}
which is finite due to Assumption \ref{assumption_1}\ref{assumption_1_L1}.
We conclude that $\| q_k^i \|_2$ is uniformly bounded for all $k \in \mathbb{Z}_{\geqslant 0}$.
Now, consider that
$
\bar{H}_{uq}(z) = \sum_{k=0}^\infty z^{-k} \begin{bmatrix} q_k^1 & \cdots & q_k^{n_p} \end{bmatrix}
$
which is analytic if the summation converges. 
Convergence is guaranteed for $|z| > 1$, due to the uniform bounds on $\|q_k^i\|_2$.
We conclude that $\bar{H}_{uq}(z)$ is analytic for $z \in \mathbb{C}_{>1}$. 
Analogous arguments hold for $\bar{H}_{ud}(z)$.
$\bar{H}_{wd}(z)$, and $\bar{H}_{wa}(z)$ are analytic in $\mathbb{C}_{>1}$ by definition, because they are spectral factors. 

To prove claim (\ref{properties_1}), it is already known that each  function is analytic for $z \in \mathbb{C}_{>1}$, and it remains to verify that the required frequency-domain norms hold.
For $\bar{H}_{uq}$, we require that 
\begin{equation*}
\| \bar{H}_{uq} \|_{\mathbb{H}_\infty} = 
\sup\limits_{\Omega \in [-\pi,\pi]} \| \bar{H}_{uq}(e^{j\Omega}) \|_2 < \infty.
\end{equation*}
Letting $\sigma_\ell \triangleq \sinc(\omega_\ell T/2)$, we have that
\begin{align*}
\| \bar{H}_{uq}(e^{j\Omega}) \|_2
=& \left\| \sum\limits_{\ell = -\infty}^\infty \hat{Z}_{uv}(j\omega_\ell) \sigma_\ell^2  \right\|_2 
\leqslant \| \hat{Z}_{uv} \|_{\mathbb{L}_\infty}  \sum\limits_{\ell = -\infty}^\infty \sigma_\ell^2
\end{align*}
But $\sum_{\ell=-\infty}^\infty \sigma_\ell^2 = 1$ so we conclude that $\| \bar{H}_{uq} \|_{\mathbb{H}_\infty} \leqslant \| \hat{Z}_{uv} \|_{\mathbb{L}_\infty}$. 
Per Assumption \ref{assumption_1}\ref{assumption_1_domains}, we conclude that $\bar{H}_{uq} \in \mathbb{H}_\infty$. 
To prove that $\bar{H}_{ud} \in \mathbb{H}_\infty$, we have that 
\begin{align*}
 \| \bar{H}_{ud}(e^{j\Omega}) \|_2
&= \left\| \frac{1}{e^{j\Omega}-1} \sum\limits_{\ell=-\infty}^\infty \hat{Z}_{uv}(j\omega_\ell) \sigma_\ell^2 \right\|_2 \\
& = \left\| \sum\limits_{\ell=-\infty}^\infty \hat{Z}_{ud}(j\omega_\ell) \left| \sigma_\ell \right| \right\|_2 \\
& \leqslant \left\| \hat{Z}_{ud}(j\omega_0) \right\|_2 |\sigma_\ell|  + \sum\limits_{\substack{\ell = -\infty \\ \neq 0 } }^\infty \left\| \hat{Z}_{uv}(j\omega_\ell) \right\|_2 \left| \frac{\sigma_\ell }{\omega_\ell} \right| 
\\
& \leqslant 
\left\| \hat{Z}_{ud} \right\|_{\mathbb{L}_\infty} + \left\| \hat{Z}_{uv} \right\|_{\mathbb{L}_\infty} \sum\limits_{\ell=1}^\infty \frac{T}{(\pi\ell)^2} 
\end{align*}
Due to Assumption \ref{assumption_1}\ref{assumption_1_domains}, the above bound is finite, implying that $\| \bar{H}_{ud} \|_{\mathbb{H}_\infty}$ is finite. 

To prove that $\bar{H}_{wa} \in \mathbb{H}_2$, equations \eqref{Sigma_q_a_1} and \eqref{Sigma_q_a_2} imply
\begin{align*}
2\pi \| \bar{H}_{wa} \|_{\mathbb{H}_2}^2 
&= \sum\limits_{\ell=-\infty}^\infty \int_{-\pi}^\pi \frac{1}{T} \sigma_\ell^2 S_{a}(\omega_\ell) d\Omega 
\leqslant \int_{-\infty}^\infty S_a(\omega) d\omega
\end{align*}
which is finite due to Assumption \ref{assumption_1}\ref{assumption_1_domains}.
A similar argument proves the same property for $\bar{H}_{wd}$. 
To prove that $\bar{H}_{wa} \in \mathbb{H}_\infty$, it suffices to show that 
$\| \bar{H}_{wa}(e^{j\Omega}) \|_2$ is uniformly bounded over $\Omega \in [-\pi,\pi]$. Again using \eqref{Sigma_q_a_1} and \eqref{Sigma_q_a_2},
\begin{align*}
\| \bar{H}_{wa}(e^{j\Omega})  \|_2^2 
\leqslant& \sum\limits_{\ell=-\infty}^\infty \frac{1}{T} \sigma_\ell^2 S_a(\omega_\ell) 
\leqslant \frac{1}{T} \| S_a \|_{\mathbb{L}_\infty}^2 
\end{align*}
which is finite due to Assumption \ref{assumption_1}\ref{assumption_1_domains}. 
A similar argument proves the same property for $\bar{H}_{wd}$.

For claim (\ref{properties_2}), observe that for $\Omega \in [-\pi,\pi]$,
\begin{equation*}
\bar{H}_{uq}(e^{j\Omega}) + \bar{H}_{uq}^H(e^{j\Omega}) 
= \sum\limits_{\ell = -\infty}^\infty \left[ \hat{Z}_{uv}(j\omega_\ell) + \hat{Z}_{uv}^H(j\omega_\ell) \right] \sigma_\ell^2
\end{equation*}
which is positive-semidefinite as per Assumption \ref{assumption_1}\ref{assumption_1_passivity}.
Because $\bar{H}_{uq}(z)$ is analytic on $\mathbb{C}_{>1}$, $\bar{H}_{uq}(z^{-1})$ is analytic on $\mathbb{C}_{< 1}$.
Define
\begin{equation*}
\bar{G}_{uq}(z) \triangleq \left[ I - \epsilon \bar{H}_{uq}(z^{-1}) \right] \left[ I + \epsilon \bar{H}_{uq}(z^{-1}) \right]^{-1} 
\end{equation*}
where $0 < \epsilon <  \| \bar{H}_{uq} \|_{\mathbb{H}_\infty}^{-1}$.
We then have that \eqref{properties_theorem_pr} holds if and only if $\| \bar{G}_{uq} (z)\| \leqslant 1$ for all $z \in \mathbb{C}_{\leqslant 1}$. 
But if $\bar{H}_{uq}(z^{-1})$ is analytic on $\mathbb{C}_{<1}$ then so is $\bar{G}_{uq}(z)$, and therefore via the Maximum Modulus Theorem, $\| \bar{G}_{uq}(z)\|$ attains is maximum on the boundary of $\mathbb{C}_{\leqslant 1}$, i.e., for $z = e^{j\Omega}$, $\Omega \in [-\pi,\pi]$. 
Because $\bar{H}_{uq}(e^{j\Omega)} + \bar{H}_{uq}(e^{-j\Omega}) \geqslant 0$, it follows that $\sup_{\Omega\in[-\pi,\pi]} \| \bar{G}_{uq}(e^{j\Omega})\| \leqslant 1$. This proves \eqref{properties_theorem_pr}. 

To prove the last caveat in claim (\ref{properties_2}), first consider the case where $|z| = 1$, i.e., $z = e^{j\Omega}$ for $\Omega \in [-\pi,\pi]$. Due to Assumption \ref{assumption_1}\ref{assumption_1_passivity}, evaluation of $\bar{H}_{uq}(e^{j\Omega})$ as in \eqref{Huq_1} gives that inequality \eqref{properties_theorem_pr} must hold strictly unless $\sinc(\omega_\ell T /2) = 0$ for all $\ell\in\mathbb{Z}\setminus \{0\}$. 
This only occurs if $\Omega =0$, i.e., $z = 1$. 
Now, consider the case in which $z \in \mathbb{C}_{>0}$. 
Let $\nu\in\mathbb{C}^{n_p}$ and define $\bar{h}(z) \triangleq \nu^H \bar{H}_{uq}(z) \nu$.
Consider the situation in which there exists a $z_0 \in \mathbb{C}_{>1}$ such that $\textrm{Re} \left\{ \bar{h}(z_0) \right\} = 0$, and define analytic function $\bar{g}(z) \triangleq \bar{h}(z) - j \textrm{Im}\{ \bar{h}(z_0) \}$.
Then $\bar{g}(z_0) = 0$.
Suppose it is the case that $\bar{g}(z) \neq 0$ for almost all $z$ in an infinitesimal neighborhood of $z = z_0$. 
Let $z = z_0 ( 1 + \delta e^{j\theta} )$ where $\delta \in \mathbb{R}_{>0}$ is infinitesimal and $\theta \in [-\pi,\pi]$. 
Because $\bar{g}(z)$ is analytic at $z = z_0$, and $\delta$ is infinitesimal,
\begin{equation*}
\bar{g}(z) = \left( z-z_0 \right)^m \psi(z_0) = \left( z_0^m \delta^m \psi(z_0) \right) e^{jm\theta}
\end{equation*}
where
\begin{equation*}
\psi(z_0) = \lim\limits_{z\rightarrow z_0} \frac{ \bar{g}(z) }{(z-z_0)^m} 
\end{equation*}
and $m\in\mathbb{Z}_{>0}$ is the lowest integer such that $\psi(z_0) \neq 0$. 
It has been proven that $\textrm{Re}\{\bar{g}(z)\} = \textrm{Re}\{\bar{h}(z)\} \geqslant 0$ for all $z \in \mathbb{C}_{>1}$ and therefore we require that 
\begin{equation}\label{properties_proof_1}
z_0^m \psi(z_0) e^{jm\theta} + z_0^{mH} \psi^H(z_0) e^{-jm\theta} \geqslant 0 , \ \ \forall \theta \in [-\pi,\pi]
\end{equation}
Because $z_0\psi(z_0) \neq 0$, this is impossible. We therefore conclude that if $\bar{g}(z_0) = 0$ for some $z_0 \in \mathbb{C}_{>1}$ then it must be the case that $\bar{g}(z) = 0$ in some open neighborhood of $z = z_0$.  
But in order for $\bar{g}(z)$ to be analytic, this implies that $\bar{g}(z) = 0$ for all $z \in \mathbb{C}_{>1}$. 
If this is true then $\bar{g}(z) = 0$ almost everywhere on the boundary of $\mathbb{C}_{>1}$, i.e., for $z = e^{j\Omega}$ with $\Omega \in [-\pi,\pi]$. 
This in turn implies that $\textrm{Re}\{\bar{h}(e^{j\Omega})\} = 0$ for almost all $\Omega \in [-\pi,\pi]$, which contradicts the last caveat in assumption \ref{assumption_1}\ref{assumption_1_passivity}. 
We therefore conclude that $\textrm{Re}\{\bar{h}(z)\} > 0$ everywhere in $\mathbb{C}_{>1}$.
This holds for all $\nu\in\mathbb{C}^{n_p}\setminus\{0\}$, implying that $\bar{H}_{uq}(z)+\bar{H}_{uq}^H(z)$ is nonsingular over the same domain.

The first part of claim (\ref{properties_6}) is immediate from \eqref{Hbar_ud}, together with claim (\ref{properties_1}).
For the second part of the claim, we make a similar argument as above, resulting in the requirement that \eqref{properties_proof_1} hold at $z_0 = 1$.
However, since $z_0 = 1$ is on the boundary of $\mathbb{C}_{\geqslant 1}$, rather than in the interior, it is only necessary that condition \eqref{properties_proof_1} hold for $\theta \in [-\tfrac{\pi}{2},\tfrac{\pi}{2}]$. 
Satisfaction of the condition requires $m = 1$ and $z_0\gamma(z_0) \in \mathbb{R}_{>0}$.
This proves that $\nu^H \left( \bar{H}_{ud}(1) + \bar{H}_{ud}^H(1) \right) \nu \in \mathbb{R}_{>0}$. 
Enforcing this for all $\nu \in \mathbb{C}^{n_p}$ implies that $ \bar{H}_{ud}(1) + \bar{H}_{ud}^H(1) > 0$.

\subsection{Proof of Theorem \ref{finite_dimensional_model_properties_theorem}}

To prove claim (\ref{Schur_property}), it is known from Theorem \ref{discrete_time_properties_theorem} that $\left\{ \bar{H}_{ud}, \bar{H}_{wa}, \bar{H}_{wd} \right\} \subset \mathbb{H}_\infty$. 
Meanwhile, the mapping $\rv{u} \mapsto \rv{a}$ is zero.
As such, mapping $\{\rv{u},\rv{w}\} \mapsto \{\rv{d},\rv{a}\}$ is in $\mathbb{H}_\infty$.
Minimality of realization \eqref{discrete_time_state_space_orig} then implies that $A$ is Schur.

To prove claim (\ref{Feedthrough_property}), it is a standard fact that $
D_{uq} =
\tfrac{1}{2\pi} \int_{-\pi}^\pi \bar{H}_{uq}(e^{j\Omega}) d\Omega
$
and consequently
\begin{equation*}
\frac{1}{2\pi} \int_{-\pi}^\pi \left[ \bar{H}_{uq}(e^{j\Omega}) + \bar{H}_{uq}^H(e^{j\Omega}) \right] d\Omega = D_{uq} + D_{uq}^T
\end{equation*}
The integrand is positive-definite for almost all $\Omega \in [-\pi,\pi]$ so the integral is positive-definite. 

To prove claim (\ref{KYP_lemma_property}), first partition the state space to isolate the subspace that is unobservable from $q$, i.e., 
\begin{align*}
A =& \begin{bmatrix} A_{11} & 0 \\ A_{21} & A_{22} \end{bmatrix}, & 
B_u =& \begin{bmatrix} B_1 \\ B_2 \end{bmatrix}, & 
C_q =& \begin{bmatrix} C_1^T \\ 0 \end{bmatrix}^T 
\end{align*}
Presume that the theorem is known to hold for the observable subspace, i.e., there exists a matrix $W_{11}=W_{11}^T$ such that 
\begin{equation*}
W_{11} = A_{11}^T W_{11} A_{11} + F_1 M F_1^T 
\end{equation*}
where 
\begin{align*}
M =& R + \tfrac{1}{2} \left( D_{uq}+D_{uq}^T \right) - B_1^T W_{11} B_1 \\
F_1 =& -M^{-1} \left[  \tfrac{1}{2} C_1  - B_1^T W_{11} A_{11} \right]
\end{align*}
Then the equations in the theorem are satisfied with $W = \textrm{blockdiag}\left\{ W_{11}, 0 \right\}$.
Henceforth we presume the state space has been reduced by eliminating the unobservable subspace.

Now assuming the reduced system to be observable, we change coordinates to isolate the controllable subspace, i.e., 
\begin{align*}
A =& \begin{bmatrix} A_{11} & A_{12} \\ 0 & A_{22} \end{bmatrix}, & 
B_u =& \begin{bmatrix} B_1 \\ 0 \end{bmatrix}, & 
C_q =& \begin{bmatrix} C_1^T \\ C_2^T \end{bmatrix}^T 
\end{align*}
where we note that $A_{11}$ and $A_{22}$ are both Schur, because $A$ is.
Partitioning $W$ similarly, we have that Riccati equation \eqref{KYP_riccati} is equivalent to the partitioned equations
\begin{align}
W_{11} =& A_{11}^T W_{11} A_{11} + F_1^T M F_1 
\label{kyp_proof_1} \\
W_{12} =& A_{11}^T W_{11} A_{12} + A_{11}^T W_{12} A_{22} + F_1^T M F_2 
\label{kyp_proof_2}\\
W_{22} =& A_{12}^T W_{11} A_{12} + A_{22}^T W_{12}^T A_{12}  \\
              &+ A_{12}^T W_{12} A_{22} + A_{22}^T W_{22} A_{22} + F_2^T M F_2
\label{kyp_proof_3}
\end{align}
where
\begin{align}
F_1 =& -M^{-1} \left[ \tfrac{1}{2} C_1 - B_1^T W_{11} A_{11} \right] \label{kyp_proof_4} \\
F_2 =& -M^{-1} \left[ \tfrac{1}{2} C_2 - B_1^T W_{11} A_{12} - B_1^T W_{12} A_{22} \right] 
\label{kyp_proof_5} \\
M =& R + \tfrac{1}{2}(D_{uq}+D_{uq}^T) - B_u^T W_{11} B_u
\label{kyp_proof_6}
\end{align}
Noting that the system 
\begin{equation*}
\left[ \begin{array}{c|c} A_{11} & B_1 \\ \hline C_1 & D_{uq} \end{array} \right]
\end{equation*}
is a minimal realization of positive-real transfer function $\bar{H}_{uq}$, the transfer function $\bar{H}'_{uq} \triangleq \bar{H}_{uq} + \tfrac{1}{2}R$ satisfies 
\begin{equation*}
\bar{H}'_{uq}(z) + \left[ \bar{H}'_{uq}(z) \right]^H \geqslant R , \ \ \forall z \in \mathbb{C}_{\geqslant 1}
\end{equation*}
Furthermore, because $A$ is Schur and $R > 0$, it follows through a perturbation argument that for $\rho \in \mathbb{R}_{>0}$ sufficiently small, 
\begin{equation*}
\bar{H}'_{uq}((1-\rho)z) + \left[ \bar{H}'_{uq}((1-\rho)z) \right]^H  > 0 , \ \ \forall z \in \mathbb{C}_{\geqslant 1}
\end{equation*}
As such, we conclude that $\bar{H}'_{uq}(z)$ is strictly positive-real (SPR). 
It then follows from the classical Kalman-Yakubovic-Popov Lemma that there exists a unique $W_{11}=W_{11}^T>0$ satisfying \eqref{kyp_proof_1}, with $F_1$ as in \eqref{kyp_proof_4} and $M>0$ as in \eqref{kyp_proof_6}, and furthermore, that $A_{11}+B_1 F_1$ is Schur.

Next, consider that \eqref{kyp_proof_2} can be written in terms of the (known) solution to $W_{11}$, as
\begin{align*}
W_{12} =& \left[ A_{11} + B_1 F_1 \right]^T W_{12} A_{22}  
\\ & + 
A_{11}^T W_{11} A_{12} - F_1^T \left[ \tfrac{1}{2} C_2 - B_1^T W_{11} A_{12} \right]
\end{align*}
This is a discrete-time Sylvester equation for $W_{12}$, which has a unique solution because $A_{11} + B_1 F_1$ and $A_{22}$ are Schur.
With $W_{12}$ solved, \eqref{kyp_proof_3} becomes a discrete-time Lyapunov equation for $W_{22}$, which has a unique solution because $A_{22}$ is Schur.

We conclude that in the case where $(A,C_q)$ is observable, a unique $W=W^T$ exists which satisfies \eqref{KYP_riccati} and renders $A+B_uF$ Schur. It remains only to show that for this observable case, $W > 0$. 
The fact that $W \geqslant 0$ is immediate from \eqref{KYP_riccati}, because $A$ is Schur and \eqref{KYP_riccati} constitutes a discrete-time Lyapunov equation for $W$ for fixed $F$ and $M$. 
To show that $W > 0$ it suffices to show that $(A,F)$ is observable. 
This can be shown by contradiction; if it were not, there would exist a vector $\eta$ and eigenvalue $\lambda$ such that $A\eta = \eta \lambda$ and $F \eta = 0$. 
But if this were the case, it would follow that from a quadratic form on \eqref{KYP_riccati} that $\eta^H W \eta \left( 1 - |\lambda|^2 \right) = 0$.
But because $A$ is Schur, it is known that $|\lambda| < 1$ so this implies that $\eta^H W \eta = 0$. 
If this is true, then because $W$ is Hermitian and positive-semidefinite, it must be the case that $W \eta = 0$.
But if this is the case, then $F\eta = -\tfrac{1}{2} M^{-1} C_q \eta$.
Via the Popov-Belevitch-Hautus test, this violates the assumption that $(A,C_q)$ is observable. 
As such, we arrive at a contradiction. 

\subsection{Proof of Theorem \ref{Schur_theorem}}

The matrix 
\begin{equation*}
A' \triangleq \begin{bmatrix} I & 0 \\ B_u^\perp A B_u [C_dB_u]^{-1} & A_s \end{bmatrix}
\end{equation*}
is similar to the matrix $A-B_uD_{uq}^{-1}C_q$.
If $\eta$ is an eigenvector of $A_s$, then $[ 0 \  \eta^H]^H$ is an eigenvector of $A'$. 
Likewise, the eigenvalues of $A'$ are those of $A_s$, together with $\lambda = 1$.
We conclude that if the eigenvalues of $A-B_uD_{uq}^{-1}C_q$ are all in $\mathbb{C}_{<1} \cup \{1\}$, then these same properties hold for $A_s$.

Without loss of generality, we assume that the system state space is partitioned to isolate the subspace controllable from $u$, i.e., we presume
\begin{align*}
A =& \begin{bmatrix} A_{11} & A_{12} \\ 0 & A_{22} \end{bmatrix}, &
B_u =& \begin{bmatrix} B_1 \\ 0 \end{bmatrix}, &
C_q =& \begin{bmatrix} C_1^T \\ C_2^T \end{bmatrix}^T
\end{align*}
and we have that 
\begin{equation*}
A-B_u D_{uq}^{-1}C_q = \begin{bmatrix} \Theta & A_{12}-B_1D_{uq}^{-1}C_2 \\ 0 & A_{22} \end{bmatrix}
\end{equation*}
where $\Theta \triangleq A_{11} - B_1D_{uq}^{-1}C_1$.  
It is known that the eigenvalues of $A-B_u D_{uq}^{-1} C_q$ are those of $\Theta$ and $A_{22}$.
Furthermore, the eigenvalues of $A_{22}$ have moduli strictly less than $1$ because $A$ is Schur.
So it follows that if $\lambda$ is an eigenvalue of $A-B_uD_{uq}^{-1}C_q$ with $|\lambda| \geqslant 1$ then it is also an eigenvalue of $\Theta$.

Now consider that 
\begin{equation*}
\bar{H}_{uq}^{-1}
\sim \left[ \begin{array}{c|c}  \Theta & B_{1}D_{uq}^{-1} \\ \hline -D_{uq}^{-1} C_{1} & D_{uq}^{-1} \end{array} \right] 
\end{equation*}
where we note that the above realization is minimal.
Consequently, $\lambda \in \mathbb{C}$ is a pole of $\bar{H}_{uq}^{-1}(z)$ if and only if it is an eigenvalue of $\Theta$.
Because $\bar{H}_{uq}(z)+\bar{H}_{uq}^H(z) > 0$ for all $z \in \mathbb{C}_{\geqslant 1} \setminus \{1\} $, it follows that $\bar{H}_{uq}(z)$ is nonsingular in this domain, and consequently $\bar{H}_{uq}^{-1}(z)$ cannot have poles there.

We conclude that all eigenvalues of $A_s$ lie in $\mathbb{C}_{<1} \cup \{1\}$.
Assume $\lambda=1$ is an eigenvalue of $A_s$. Then there exists a vector $\eta$ such that
\begin{align*}
0 =& \eta^H \left[ A_s - I \right] 
= \eta^H B_u^\perp \left[ A - I \right] C_d^\perp [ B_u^\perp C_d^\perp]^{-1}
\nonumber
\end{align*}
implying that there exists a vector $\nu$ such that
\begin{equation*}
 \eta^H B_u^\perp \left[ A - I \right]
 = \nu^H C_d.
\end{equation*}
Because $A$ is known to be Schur, it follows that $A-I$ is nonsingular, and consequently $\nu$ must be nonzero.
Multiplying from the right by $[A-I]^{-1}B_u$ gives that 
$\nu^H \bar{H}_{ud}(1) = 0$.
It follows that $\lambda=1$ is a zero of scalar transfer function $\nu^H \bar{H}_{ud}(z) \nu$.  But 
$\bar{H}_{uq}(z) = \tfrac{1}{T} (z-1) \bar{H}_{ud}(z)$
so we conclude that $\lambda=1$ must also be a zero of scalar transfer function $\nu^H \bar{H}_{uq}(z) \nu$.
If $\nu^H \bar{H}_{ud}(z) \nu$ has a zero at $z = 1$, then $\nu^H \bar{H}_{uq}(z) \nu$ must have a repeated zero at $z = 1$.
But because $\bar{H}_{uq}(z)$ is positive-real, it follows that $\nu^H \bar{H}_{uq}(z) \nu$ is positive-real, precluding the possibility of a repeated zero at $z = 1$.
As such, we arrive at a contradiction, and thus conclude that $\lambda = 1$ cannot be an eigenvalue of $A_s$. 

\subsection{Proof of Theorem \ref{optimal_control_theorem}}

We have that for some time $k \in \mathbb{Z}$,
\begin{align*}
\Ex\{ \rv{p}_k \} =& -\mathcal{E} \left\{ \rv{u}^T_k C_q \rv{x}_k + \| \rv{u}_k \|_{\tilde{R}}^2 + v_d^T |\rv{u}_k| \right\}
\end{align*}
where we recall that $\tilde{R} = \tilde{R}^T > 0$.
It is straight-forward to verify that 
\begin{align*}
\mathcal{E} \left\{  \| \rv{u}_k \|_{\tilde{R}}^2  \right\} 
=& \mathcal{E} \left\{ \| \rv{\xi}_k \|_M^2 \right\} + \mathcal{E} \left\{ \| A\rv{x}_k+B_u\rv{u}_k \|_W^2 \right\} \\
&- \mathcal{E} \left\{ \| \rv{x}_k \|_W^2 \right\} - \mathcal{E}\left\{ \rv{u}_k^T C_q \rv{x}_k \right\}
\end{align*}
Because $\rv{x}_k$ and $\rv{u}_k$ are independent of $\rv{w}_k$, and because $\rv{w}_k$ is zero-mean with $\mathcal{E}\{\rv{w}_k\rv{w}_k^T\} = S_w$, 
\begin{align*}
\mathcal{E} \left\{  \| \rv{u}_k \|_{\tilde{R}}^2  \right\} 
=& \mathcal{E} \left\{ \| \rv{\xi}_k \|_M^2 \right\} - \mathcal{E}\left\{ \rv{u}_k^T C_q \rv{x}_k \right\} - \bar{p} \\
& + \mathcal{E} \left\{ \| \rv{x}_{k+1} \|_W^2 \right\} - \mathcal{E} \left\{ \| \rv{x}_k \|_W^2 \right\}.
\end{align*}
Summing over $k \in \{0,...,\tau\}$ and dividing by $\tau+1$ gives the claimed result, if \eqref{optimal_control_theorem_eq1} holds.

\subsection{Proof of Lemma \ref{Jf_lemma}}

Substitute 
\eqref{stall_state_transition}, \eqref{stall_control_input}, and \eqref{stall_slack_variable} into \eqref{rho}, with $J(\cdot,\cdot,\cdot)$ defined as in \eqref{J}, to get
\begin{align} 
&\rho(x_{k+h+1|k}) \nonumber \\
&= x_{k+h+1|k}^T \left( \sum\limits_{i=0}^\infty Q_i \right)  x_{k+h+1|k} 
+ v_d^T \left( \sum\limits_{i=0}^\infty \left| u_{k+h+1+i|k} \right| \right) \nonumber \\
& \quad + \mu \left( \sum\limits_{i=0}^\infty u_{\max}^T \max \left\{ 0 , \left| u_{k+h+1+i|k} \right| - u_{\max} \right\} \right)
\label{Vtilde_2}
\end{align}
where, for $i \in \mathbb{Z}_{\geqslant 0}$,
\begin{align*}
Q_i \triangleq & (B_u^\perp)^T (A_s^i)^T C_s^T (F+D_{uq}^{-1}C_q)^T \nonumber \\ & \quad\quad \times 
M (F+D_{uq}^{-1}C_q) C_s A_s^i B_u^\perp 
\end{align*}
By Theorem \ref{Schur_theorem}, $A_s$ is Schur, and so
$
\sum_{i=0}^\infty Q_i = (B_u^\perp)^T Z B_u^\perp
$
where $Z=Z^T \geqslant 0$ is the unique solution to the discrete-time Lyapunov equation
\begin{equation*}
Z = A_s^T Z A_s + C_s^T (F+D_{uq}^{-1}C_q)^T M (F+D_{uq}^{-1}C_q) C_s
\end{equation*}

There is no convenient closed-form solution for the second summation in \eqref{Vtilde_2}, so we seek to over-bound it by a simple function of $x_{k+h+1|k}$. 
To do this, first let $\Pi \in \mathbb{C}^{n\times n}$ be a similarity transformation that converts $A_s$ to its Jordan form $\Lambda$, i.e., $A_s = \Pi \Lambda \Pi^{-1}$.  
Then we have that
\begin{align*}
\left| u_{k+h+1+i|k} \right| =& \sum\limits_{i=0}^\infty \left| D_{uq}^{-1} C_q C_s \Pi \ \Lambda^i  \  \Pi^{-1} \ B_u^\perp x_{k+h+1|k} \right| \\
\leqslant & \sum\limits_{i=0}^\infty \left| D_{uq}^{-1} C_q C_s \Pi \right| \ \left| \Lambda \right|^i \ \left| \Pi^{-1} B_u^\perp x_{k+h+1|k} \right| \\
=& \left| D_{uq}^{-1} C_q C_s \Pi \right| \left[ I - \left| \Lambda \right| \right]^{-1} \left| \Pi^{-1} B_u^\perp x_{k+h+1|k} \right| 
\end{align*}
where we note that the above infinite summation converges because if $\Lambda$ is a Schur Jordan form, then so is $|\Lambda|$. 

The last term in \eqref{Vtilde_2} can be conservatively over-bounded by recognizing that 
\begin{align}\label{force_penalty_bound}
u_{\max}^T \max\{ 0 , |u_{k+h+1+i|k}| - u_{\max} \} 
\leqslant & \tfrac{1}{4} \| u_{k+h+1+i|k} \|_2^2
\end{align}
But
\begin{align*}
\sum\limits_{i=0}^\infty \| u_{k+h+1+i|k} \|^2 
=& x_{k+h+1|k}^T (B_u^\perp)^T X B_u^\perp x_{k+h+1|k}
\end{align*}
where $X=X^T\geqslant 0$ is the solution to Lyapunov equation
\begin{equation*}
X = A_s^T X A_s + C_s^T C_q^T D_{uq}^{-T} D_{uq}^{-1} C_q C_s.
\end{equation*}
We arrive at an over-bound for $\rho(x_{k+h+1|k})$ as in \eqref{Jf_lemma_ineq1}, with
\begin{align*}
Q_f =& (B_u^\perp)^T (Z+\tfrac{1}{4}\mu X) B_u^\perp \\
C_f =& \diag\left\{ \left[I-|\Lambda|\right]^{-T} \left| D_{uq}^{-1} C_q C_s \Pi \right|^T v_d \right\} \Pi^T B_u^\perp
\end{align*}

To prove inequality \eqref{Jf_lemma_ineq2}, consider that because $x_{k+h+2|k} = C_s A_s B_u^\perp x_{k+h+1|k}$ and $B_u^\perp C_s = I$,  we have that
\begin{align*}
J_f(x_{k+h+2|k})
=&\left\| B_u^\perp x_{k+h+1|k} \right\|_{\Xi}^2 + \| C_f x_{k+h+1|k} \|_1
\end{align*}
where
\begin{align*}
\Xi \triangleq & A_s^T (Z+\tfrac{1}{4}\mu X) A_s \\
=& (Z+\tfrac{1}{4}\mu X) - C_s^T(F+D_{uq}^{-1}C_q)^T M (F+D_{uq}^{-1}C_q) C_s \\
& \quad - \tfrac{1}{4}\mu C_s^T C_q^T D_{uq}^T D_{uq}^{-1} C_q C_s
\end{align*}
and therefore
\begin{align*}
\left\| B_u^\perp x_{k+h+1|k} \right\|_{\Xi}^2 
& = \left\| x_{k+h+1|k} \right\|_{Q_f}^2 - \tfrac{1}{4} \mu \left\| u_{k+h+1|k} \right\|^2  \\
& \quad -  \left\| u_{k+h+1|k} - F x_{k+h+1|k} \right\|_M^2
\end{align*}
Also, consider that 
\begin{align*}
&\| C_f x_{k+h+2|k} \|_1 \\
&= v_d^T \left| D_{uq}^{-1} C_q C_s \Pi \right| \left[I-|\Lambda|\right]^{-1} \left| \Pi^{-1} B_u^\perp x_{k+h+2|k} \right| \\
&= v_d^T \left| D_{uq}^{-1} C_q C_s \Pi \right| \left[I-|\Lambda|\right]^{-1} \left| \Pi^{-1} A_s B_u^\perp x_{k+h+1|k} \right| \\
&= v_d^T \left| D_{uq}^{-1} C_q C_s \Pi \right| \left[I-|\Lambda|\right]^{-1} \left| \Lambda \Pi^{-1} B_u^\perp x_{k+h+1|k} \right| \\
&\leqslant v_d^T \left| D_{uq}^{-1} C_q C_s \Pi \right| \left[I-|\Lambda|\right]^{-1} | \Lambda | \left| \Pi^{-1} B_u^\perp x_{k+h+1|k} \right| \\
&= \| C_f x_{k+h+1|k} \|_1 \\
& \quad - v_d^T \left| D_{uq}^{-1} C_q C_s \Pi \right| \left| \Pi^{-1} B_u^\perp x_{k+h+1|k} \right| \\
&\leqslant \| C_f x_{k+h+1|k} \|_1 - v_d^T \left| D_{uq}^{-1} C_q x_{k+h+1|k} \right| \\
&= \| C_f x_{k+h+1|k} \|_1 - v_d^T \left| u_{k+h+1|k} \right|
\end{align*}
Recalling \eqref{force_penalty_bound}, we conclude \eqref{Jf_lemma_ineq2}. 

\subsection{Proof of Theorem \ref{stability_theorem}}

We begin with two technical lemmas which establish various bounds that will be useful in the sequel.

\begin{lm1} \label{bounding_lemma_1}
Let $K : Y_k \mapsto u_k$ be as in \eqref{MPC_algorithm_1}.
Then there exist constants $\{c_1,c_2\} \in \mathbb{R}_{> 0}$ such that 
\begin{equation}
\Gamma(x_k,\mathring{U}_{k|k},\mathring{B}_{k|k}) \leqslant c_2 \| x_k \|_2^2 + c_1 \| x_k \|_1
\end{equation}
\end{lm1}
\begin{proof}
For a given $x_k$, let the (sub-optimal) sequence $U_{k|k}$ be the sequence rendering $d_{m|k} = 0$ for all $m > k$, i.e., 
\begin{equation*}
u_{m|k} = \left\{ \begin{array}{lll} 
-\tfrac{1}{T} D_{uq}^{-1} C_d A x_k &:& m = k \\ 
-D_{uq}^{-1}C_qC_sA_s^{m-k} B_u^\perp x_k &:& m > k
\end{array} \right.
\end{equation*}
and let corresponding sequence $B_{k|k}$ be the minimum value that satisfies the force constraint, i.e., 
\begin{equation*}
b_{m|k} = \max\{ 0, |u_{m|k}|-u_{\max} \}.
\end{equation*}
Then the fact that $\{U_{k|k},B_{k|k}\}$ are feasible for initial condition $x_k$ implies that that $\Gamma(x_k,U_{k|k},B_{k|k}) \geqslant \Gamma(x_k,\mathring{U}_{k|k},\mathring{B}_{k|k})$.  Furthermore, by repeated application of Lemma \ref{Jf_lemma} that
\begin{align*}
&\Gamma(x_k,U_{k|k},B_{k|k}) \\
&\leqslant J_f(x_{k+1|k}) + J(x_k,u_{k|k},b_{k|k}) \\
&= \| x_{k+1|k} \|_{Q_f}^2 + \| C_f x_{k+1|k} \|_1 
 + \| u_{k|k}-Fx_{k|k} \|_M^2 \\ & \quad + (v_d+\mu u_{\max})^T | u_{k|k} |  
\end{align*}
where we have used the fact that $b_{k|k} \leqslant |u_{k|k}|$. 
Substitution of the expression above for $u_{k|k}$, and using the fact that $C_s B_u^\perp = I$, we have that 
\begin{align*}
&\Gamma(x_k,U_{k|k},B_{k|k}) \leqslant 
\| \check{A} x_k \|_{Q_f}^2  + \| C_f \check{A} x_k \|_1 \\
& \quad + \| (F+\tfrac{1}{T}D_{uq}^{-1}C_d) x_k \|_M^2  + (v_d+\mu u_{\max})^T | \tfrac{1}{T} D_{uq}^{-1} C_d x_k |  
\end{align*}
where $\check{A} \triangleq (I-\tfrac{1}{T}B_uD_{uq}^{-1}C_d)A$.
It is then straight-forward to show by defining $c_1$ and $c_2$ as
\begin{align*}
c_1 =& \| C_f \check{A} \|_1 + \left\| \diag\{v_d+\mu u_{\max} \}\tfrac{1}{T} D_{uq}^{-1} C_d \right \|_1 \\
c_2 =& \| Q_f^{1/2} \check{A} \|_2^2 + \| M^{1/2} (F+\tfrac{1}{T}D_{uq}^{-1}C_d ) \|_2^2 
\end{align*}
we have that
\begin{equation*}
\Gamma(x_k,U_{k|k},B_{k|k}) \leqslant  c_2 \| x_k \|_2^2 + c_1 \| x_k \|_1.
\end{equation*}
Recalling that $\Gamma(x_k,U_{k|k},B_{k|k}) \geqslant \Gamma(x_k,\mathring{U}_{k|k},\mathring{B}_{k|k})$ completes the proof.
\end{proof}

\begin{lm1}\label{bounding_lemma_2}
For two initial states $x_k^i$, $i \in \{1,2\}$, let $U_{k|k}^i$ and $B_{k|k}^i$ be feasible sequences for MPC control algorithm \eqref{MPC_algorithm_1}, i.e., trajectories which satisfy
\begin{align}
\Theta(x_k^i,U_{k|k}^i,B_{k|k}^i) \leqslant & 0,  &
\Psi(x_k^i,U_{k|k}^i) =& 0 
\end{align}
Let the corresponding receding-horizon trajectories with these input sequences be $x_{m|k}^i$ for $m \geqslant k$.
Define perturbations $\tilde{U}_{k|k} = U_{k|k}^2 - U_{k|k}^1$ and $\tilde{B}_{k|k} = B_{k|k}^2 - B_{k|k}^1$, and $\tilde{x}_{m|k} = x_{m|k}^2 - x_{m|k}^1$.  
Then 
\begin{align}
& \Gamma(x_k^2,U_{k|k}^2,B_{k|k}^2) \leqslant \Gamma(x_k^1,U_{k|k}^1,B_{k|k}^1) \nonumber \\ 
& \quad + 2 \Gamma^{1/2} (x_k^1,U_{k|k}^1,B_{k|k}^1) \chi_k^{1/2}
 + \Gamma(\tilde{x}_k,\tilde{U}_{k|k},|\tilde{B}_{k|k}|)
 \label{Gamma_perturbation_ineq}
\end{align}
where
\begin{equation}
\chi_k \triangleq \sum\limits_{m=k}^{k+h} \| \tilde{u}_{m|k}-F\tilde{x}_{m|k} \|_M^2 + \| \tilde{x}_{k+h+1|k} \|_{Q_f}^2
\end{equation}
\end{lm1}
\begin{proof}
We have that
\begin{align*}
& \Gamma(x_k^2,U_{k|k}^2,B_{k|k}^2) \\
& =  
\sum\limits_{m=k}^{k+h} \left\| u_{m|k}^1 - F x_{m|k}^1 \right\|_M^2 + \sum\limits_{m=k}^{k+h}  \left\| \tilde{u}_{m|k} - F\tilde{x}_{m|k} \right\|_M^2 \\
&\quad + 2 \sum\limits_{m=k}^{k+h} \left( u_{m|k}^1-Fx_{m|k}^1 \right)^T M \left( \tilde{u}_{m|k}-F\tilde{x}_{m|k} \right) \\
&\quad + \sum\limits_{m=k}^{k+h} \left( v_d^T \left| u_{m|k}^1 + \tilde{u}_{m|k} \right| + \mu u_{\max}^T (b_{m|k}^1+\tilde{b}_{m|k}) \right) \\
&\quad + \| x_{k+h+1}^1 \|_{Q_f}^2  + \| \tilde{x}_{k+h+1} \|_{Q_f}^2 + 2 \tilde{x}_{k+h+1}^T Q_f x_{k+h+1}^1 \\
&\quad + \| C_f x_{k+h+1} + C_f \tilde{x}_{k+h+1} \|_1
\end{align*}
Via the Cauchy-Schwartz inequality,
\begin{align*}
&  \sum\limits_{m=k}^{k+h} \left( u_{m|k}^1-Fx_{m|k}^1 \right)^T M \left( \tilde{u}_{m|k}-F\tilde{x}_{m|k} \right) \\
& + \tilde{x}_{k+h+1}^T Q_f x_{k+h+1}^1 \\
& \leqslant \left( \sum\limits_{m=k}^{k+h} \left\| u_{m|k}^1-Fx_{m|k}^1 \right\|_M^2 + \| x_{k+h+1}^1 \|_{Q_f}^2 \right)^{1/2} \chi_k \\
& \leqslant \Gamma^{1/2}(x_k^1,U_{k|k}^1,B_{k|k}^1) \chi_k
\end{align*}
and via the triangle inequality,
\begin{align*}
&\sum\limits_{m=k}^{k+h} v_d^T \left| u_{m|k}^1 + \tilde{u}_{m|k} \right| + \left\| C_f x_{k+h+1}^1 + C_f \tilde{x}_{k+h+1} \right\|_1 \\
& \leqslant \sum\limits_{m=k}^{k+h} v_d^T \left| u_{m|k}^1 \right| 
+ \sum\limits_{m=k}^{k+h} v_d^T \left| \tilde{u}_{m|k} \right|
\\
&\quad + \left\| C_f x_{k+h+1}^1 \right\|_1  + \left\| C_f \tilde{x}_{k+h+1} \right\|_1 
\end{align*}
Also, note that because $b_{m|k}^1$ and $b_{m|k}^2$ are both nonnegative,
\begin{equation*}
u_{\max}^T (b_{m|k}^1 + \tilde{b}_{m|k}) \leqslant u_{\max}^T b_{m|k}^1 + u_{\max}^T |\tilde{b}_{m|k}|.
\end{equation*}
Substituting these inequalities gives \eqref{Gamma_perturbation_ineq}. 
\end{proof}

Let the present time be $k$, and let $\mathring{U}_{k|k}$ and $\mathring{B}_{k|k}$ be the trajectories found by the MPC algorithm \eqref{MPC_algorithm_1}, given present state $x_k$.
For $m \geqslant k$, let the optimized receding-horizon state trajectory be denoted $\mathring{x}_{m|k}$, with $\mathring{x}_{k|k} = x_k$.
Let $\mathring{d}_{m|k}$ be as in \eqref{dm|k}, evaluated with the optimal trajectories, i.e., $\mathring{d}_{m|k} = C_d \mathring{x}_{m|k}$, where we note that $\mathring{d}_{k+h+1|k} = 0$.
For $m > k+h$, we denote the post-horizon input $\mathring{u}_{m|k}$ as the input that maintains $\mathring{d}_{m|k} = 0$, i.e., 
\begin{align*}
\mathring{x}_{m|k} =& C_s A_s^{m-(k+h+1)} B_u^\perp \mathring{x}_{k+h+1|k} \\
\mathring{u}_{m|k} =& -D_{uq}^{-1} C_q \mathring{x}_{m|k}  
\end{align*}

Now, we advance the present time to $k+1$. 
With acquisition of data $y_k$, the data $w_k$ and $x_{k+1}$ become known. 
We consider the receding-horizon control trajectory $U_{k+1|k+1}$ that enforces
\begin{equation}\label{d=mathring(d)}
d_{m|k+1} = \mathring{d}_{m|k}, \ \ \forall m \in \{ k+2, ..., k+h+2 \}
\end{equation}
We formulate the corresponding receding-horizon slack variable trajectory $B_{k+1|k+1}$ as
\begin{equation*}
b_{m|k+1} = \max\left\{ 0 , |u_{m|k+1}| - u_{\max} \right\}
\end{equation*}
Note that, so formulated, this trajectory pair $\{ U_{k+1|k+1}, B_{k+1|k+1} \}$ satisfies the constraints in the MPC algorithm at the $(k+1)^{th}$ iteration, i.e., conditions
\begin{align*}
\Theta(x_{k+1},U_{k+1|k+1},B_{k+1|k+1}) \leqslant & 0, & 
\Psi(x_{k+1},U_{k+1|k+1}) = & 0 
\end{align*}
and is therefore in the feasibility domain for the MPC optimization algorithm at time $k+1$.

We now claim that the sequence $U_{k+1|k+1}$ resulting in \eqref{d=mathring(d)}, and the associated receding-horizon sequence $\{ x_{m|k+1} : m \geqslant k+1 \}$, can be expressed as
\begin{align}
u_{m|k+1} =& \mathring{u}_{m|k} + \tilde{u}_{m|k+1} \\
x_{m|k+1} =& \mathring{x}_{m|k} + \tilde{x}_{m|k+1}
\end{align}
where 
\begin{align}
\tilde{u}_{m|k+1} 
 &= -\tfrac{1}{T} D_{uq}^{-1} C_d A \tilde{x}_{m|k+1}
\label{ms_stability_lemma_1}
\\
\tilde{x}_{m|k+1} &= \left\{ \begin{array}{lll}
B_w w_k &:& m = k+1 \\
\tilde{A}^{m-k-1} B_w w_k &:& m > k+1 
\end{array}\right.
\label{ms_stability_lemma1_proof_1}
\end{align}
with $\tilde{A} \triangleq C_d^{\perp} [B_u^\perp C_d^\perp]^{-1} B_u^\perp A$.
To show this, consider the inverse system dynamic model from Section \ref{stall_section}. 
Let 
\begin{align*}
\zeta_{m|k+1} \triangleq & \mathcal{E}\{ \rv{\zeta}_m | \rv{Y}_{k+1}=Y_{k+1}, \ \rv{d}_m = d_{m|k+1}, \forall m \geqslant k+1 \}.
\end{align*}
Then we have that for $m \geqslant k+1$, 
\begin{align*}
\zeta_{m+1|k+1} =& A_s \zeta_{m|k+1} + B_{ds} d_{m|k+1} 
\end{align*}
with initial condition 
\begin{equation*}
\zeta_{k+1|k+1} = A_s \zeta_k + B_{ws} w_k + B_{ds} d_k
\end{equation*}
But we also have that for $m \geqslant k+1$, 
\begin{align*}
\mathring{\zeta}_{m+1|k} =& A_s \mathring{\zeta}_{m|k} + B_{ds} \mathring{d}_{m|k} \end{align*}
with initial condition
\begin{equation*}
\mathring{\zeta}_{k+1|k} = A_s \zeta_k + B_{ds} \mathring{d}_{k|k}
\end{equation*}
So, letting $\tilde{\zeta}_{m|k+1} \triangleq \zeta_{m|k+1} - \mathring{\zeta}_{m|k}$, we have that 
\begin{equation*}
\tilde{\zeta}_{m+1|k+1} =A_s \tilde{\zeta}_{m|k+1} + B_{ds} \left( d_{m|k+1} - \mathring{d}_{m|k} \right)
\end{equation*}
with initial condition
\begin{equation*}
\tilde{\zeta}_{k+1|k+1} = B_{ws} w_k + B_{ds} \left( d_k - \mathring{d}_{k|k} \right) 
\end{equation*}
But $d_k - \mathring{d}_{k|k} = D_{wd} w_k$, so using \eqref{Bs}, the above initial condition simplifies to 
\begin{equation*}
\tilde{\zeta}_{k+1|k+1} = B_{u}^\perp B_w w_k 
\end{equation*}
Also, we have that 
$d_{k+1|k+1} - \mathring{d}_{k+1|k}= C_d B_w w_k$
while for $m > k+1$, \eqref{d=mathring(d)} implies that 
$d_{m|k+1} - \mathring{d}_{m|k} = 0.$
Consequently, we have that for $m > k+1$,
\begin{align*}
\tilde{\zeta}_{m|k+1} =& A_s^{m-k-2} \left[ A_s B_u^\perp + B_{ds} C_d \right] B_w w_k
\\
=& A_s^{m-k-2} B_u^\perp A B_w w_k
\end{align*}
where we have used \eqref{As} and \eqref{Bs} in the second line.
For $m \geqslant k+1$, it follows from \eqref{stall_state_system_2} and the above, that 
\begin{align*}
\tilde{x}_{m|k+1} 
=& \left\{ \begin{array}{lll}
C_s B_u^\perp B_w w_k + D_{ds} C_d B_w w_k &:& m = k+1 \\
C_s A_s^{m-k-2} B_u^\perp A B_w w_k &:& m > k+1
\end{array} \right. 
\end{align*}
which simplifies to \eqref{ms_stability_lemma1_proof_1}.
Using \eqref{inverse_system_u_eq1}, this implies that
\begin{align*}
&\tilde{u}_{m|k+1} = -D_{uq}^{-1} C_q \tilde{x}_{m|k+1} + \tfrac{1}{T} D_{uq}^{-1} \left( \tilde{d}_{m+1|k+1} - \tilde{d}_{m|k+1} \right).
\end{align*}
Substitution of \eqref{ms_stability_lemma1_proof_1} and use of \eqref{As}-\eqref{Ds} give the above as equivalent to \eqref{ms_stability_lemma_1}. 

Define sequences $\mathring{U}_{k+1|k} = \{ \mathring{u}_{k+1|k},...,\mathring{u}_{k+h+1|k} \}$ and $\mathring{B}_{k+1|k} = \{ \mathring{b}_{k+1|k},...,\mathring{b}_{k+h+1|k} \}$.
Then we have that the pair $\{\mathring{U}_{k+1|k},\mathring{B}_{k+1|k}\}$ are in the feasibility domain of MPC optimization algorithm \eqref{MPC_algorithm_1} at time $k+1$ given initial condition $\mathring{x}_{k+1|k}$. 
From Lemma \ref{bounding_lemma_2}, we infer that
\begin{align*}
&\Gamma(x_{k+1},U_{k+1|k+1},B_{k+1|k+1}) \nonumber \\
&\leqslant \Gamma(\mathring{x}_{k+1|k},\mathring{U}_{k+1|k},\mathring{B}_{k+1|k}) \\
& \quad + 2\Gamma^{1/2} (\mathring{x}_{k+1|k},\mathring{U}_{k+1|k},\mathring{B}_{k+1|k}) \chi_{k+1} \\
& \quad + \Gamma(\tilde{x}_{k+1},\tilde{U}_{k+1|k+1},|\tilde{U}_{k+1|k+1}|) 
\end{align*}
where
\begin{equation} \label{chi_k+1}
\chi_{k+1} \triangleq \sum\limits_{m=k+1}^{k+h+1} \| \tilde{u}_{m|k+1}-F\tilde{x}_{m|k+1} \|_M^2 + \| \tilde{x}_{k+h+2|k+1} \|_{Q_f}^2
\end{equation}
From Lemma \ref{Jf_lemma} we have inequality \eqref{Jf_lemma_ineq2}, and therefore 
\begin{equation*}
\Gamma(x_k,\mathring{U}_{k|k},\mathring{B}_{k|k}) 
\geqslant J(x_k,u_k,b_k) + \Gamma(\mathring{x}_{k+1|k},\mathring{U}_{k+1|k},\mathring{B}_{k+1|k}).
\end{equation*}
So it follows that 
\begin{multline*}
\Gamma( x_{k+1} , U_{k+1|k+1}, B_{k+1|k+1} ) 
\leqslant \Gamma( x_k, \mathring{U}_{k|k}, \mathring{B}_{k|k} ) \\
- J(x_k,u_k,b_k) + 2\Gamma^{1/2} (\mathring{x}_{k+1|k},\mathring{U}_{k+1|k},\mathring{B}_{k+1|k}) \chi_{k+1} \\
 + \Gamma(\tilde{x}_k,\tilde{U}_{k|k},|\tilde{U}_{k|k}|) 
\end{multline*}
Now, using Lemma \ref{bounding_lemma_1}, we note that
\begin{align*}
&\Gamma^{1/2}(\mathring{x}_{k+1|k},\mathring{U}_{k+1|k},\mathring{B}_{k+1|k}) \\
&\leqslant \left[ \Gamma(\mathring{x}_{k|k},\mathring{U}_{k|k},\mathring{B}_{k|k}) - J(x_k,u_k,b_k) \right]^{1/2} \\
& \leqslant \Gamma^{1/2} (\mathring{x}_{k|k},\mathring{U}_{k|k},\mathring{B}_{k|k}) \\
& \leqslant \left[ c_2 \| x_k \|_2^2 + c_1 \| x_k \|_1 \right]^{1/2} \\
& \leqslant \left[ c_2 \| x_k \|_2^2 + c_1 \sqrt{n} \| x_k \|_2 \right]^{1/2} \\
& \leqslant \sqrt{c_2} \left[ \| x_k \|_2 + \frac{c_1 \sqrt{n}}{2c_2} \right]
\end{align*}
So we conclude that
\begin{multline*}
\Gamma( x_{k+1} , U_{k+1|k+1}, B_{k+1|k+1} ) 
\leqslant \Gamma( x_k, \mathring{U}_{k|k}, \mathring{B}_{k|k} ) \\
- J(x_k,u_k,b_k) + 2\sqrt{c_2} \left[ \| x_k \|_2 + \frac{c_1 \sqrt{n}}{2c_2} \right] \chi_{k+1} \\
 + \Gamma(\tilde{x}_k,\tilde{U}_{k|k},|\tilde{U}_{k|k}|) 
\end{multline*}
Now, we recognize that due to \eqref{ms_stability_lemma_1} and \eqref{ms_stability_lemma1_proof_1}, together with \eqref{Gamma_def} and \eqref{chi_k+1}, it follows that there exist constants $a_1$, $a_2$, and $a_3$ such that
\begin{align*}
& \chi_{k+1}^{1/2} \leqslant a_1 \| w_k \|_2 \leqslant a_1 \| w_k \|_1 \\
& \Gamma(\tilde{x}_k,\tilde{U}_{k|k},|\tilde{U}|_{k|k}) \leqslant a_2 \| w_k \|_2^2 + a_3 \| w_k \|_1
\end{align*}
As such, we conclude that
\begin{align*}
&J(x_k,u_k,b_k) \\
& \leqslant \Gamma(x_k,\mathring{U}_{k|k},\mathring{B}_{k|k}) - \Gamma(x_{k+1},U_{k+1|k+1},B_{k+1|k+1}) \\
& \quad 
+ a_2 \| w_k \|_2^2 + a_4 \| w_k \|_1 + a_5 \|x_k\|_2 \| w_k \|_2 
\end{align*}
where $a_4 \triangleq a_3 + a_1 c_1 \sqrt{n/c_2}$ and $a_5 \triangleq 2\sqrt{c_2}a_1$.

Now consider the implementation of MPC optimization algorithm \eqref{MPC_algorithm_1} at time $k+1$, resulting in optimal sequences $\mathring{U}_{k+1|k+1}$ and $\mathring{B}_{k+1|k+1}$.
Because sequences $U_{k+1|k+1}$ and $B_{k+1|k+1}$ derived above are feasible, it follows that 
\begin{multline*}
\Gamma(x_{k+1},\mathring{U}_{k+1|k+1},\mathring{B}_{k+1|k+1}) \\
\leqslant 
\Gamma(x_{k+1},U_{k+1|k+1},B_{k+1|k+1}) 
\end{multline*}
Define $r_\tau$ for all $\tau \in \mathbb{Z}_{\geqslant 0}$ as
\begin{equation*}
r_\tau \triangleq \sum\limits_{k=0}^\tau J(x_k,u_k,b_k)
\end{equation*}
Then the above inequalities imply that
\begin{align*}
r_\tau 
\leqslant &
\Gamma(x_0,\mathring{U}_{0|0},\mathring{B}_{0|0}) + a_2 \|w\|_{2\tau}^2 + a_4 \| w \|_{1\tau} \\
& + a_5 \sum\limits_{k=0}^{\tau}  \|x_k\|_2 \|w_k\|_2 
\end{align*}
Using Lemma \ref{bounding_lemma_1} and the Cauchy-Schwartz inequality, 
\begin{align*}
r_\tau \leqslant & c_2 \| x_0 \|_2^2 + c_1 \| x_0 \|_1 + a_2 \| w \|_{2\tau}^2 + a_4 \|w\|_{1\tau}  \\
& + a_5 \| x \|_{2\tau} \|w\|_{2\tau} 
\end{align*}

Let $\delta_k \triangleq M^{1/2} ( u_k - F x_k )$.  Then it follows that 
$r_\tau \geqslant \| \delta \|_{2\tau}^2$, and thus we have that
\begin{align*}
\| \delta \|_{2\tau}^2 \leqslant & c_2 \| x_0 \|_2^2 + c_1 \| x_0 \|_1 + a_2 \| w \|_{2\tau}^2 + a_4 \|w\|_{1\tau}  \\
& + a_5 \| x \|_{2\tau} \|w\|_{2\tau}
\end{align*}
Meanwhile, consider that 
\begin{equation*}
x_{k+1} = \left[ A + B_u F \right] x_k + B_u M^{-1/2} \delta_k + B_w w_k
\end{equation*}
where we recall from Theorem \ref{finite_dimensional_model_properties_theorem} that $[A+B_uF]$ is Schur.  
It follows that there exist $\{\vartheta, \varphi\} \subset \mathbb{R}_{>0}$ such that 
\begin{equation*}
\| x \|_{2\tau}^2 \leqslant \vartheta \| \delta \|_{2\tau}^2 + \varphi \| w \|_{2\tau}^2 
\end{equation*}
for all $\tau \in \mathbb{R}_{>0}$.  So we conclude that 
\begin{align*}
0 & \geqslant \| x \|_{2\tau}^2 - \vartheta a_5 \| w \|_{2\tau}  \| x \|_{2\tau}   \\ & - \left( \vartheta c_2 \| x_0 \|_2^2 + \vartheta c_1 \| x_0 \|_1 + (\vartheta a_2 + \varphi) \| w \|_{2\tau}^2 + \vartheta a_4 \|w\|_{1\tau} \right)
\end{align*}
The right-hand side is convex in $\|x\|_{2\tau}$, with real roots. So 
\begin{multline*}
\| x \|_{2\tau} \leqslant 
\tfrac{1}{2} \vartheta a_5 \|w\|_{2\tau} \\
+ \sqrt{ 
\begin{array}{l}
\vartheta c_2 \| x_0 \|_2^2 + \vartheta c_1 \| x_0 \|_1 + \vartheta a_4 \|w\|_{1\tau} \\
+ (\vartheta a_2 + \varphi + \tfrac{1}{4}\vartheta^2a_5^2) \| w \|_{2\tau}^2 
\end{array}
}
\end{multline*}
Inequality \eqref{stability_theorem_bound} is obtained by squaring the above expression and linearizing the (concave) square-root term about the value $\tfrac{1}{2}\vartheta a_5 \|w\|_{2\tau}$. 

\bibliographystyle{IEEEtran}
\bibliography{wec_mpc}

\vspace{-25pt}
\begin{IEEEbiography}[{\includegraphics[width=1in,height=1.25in,clip,keepaspectratio]{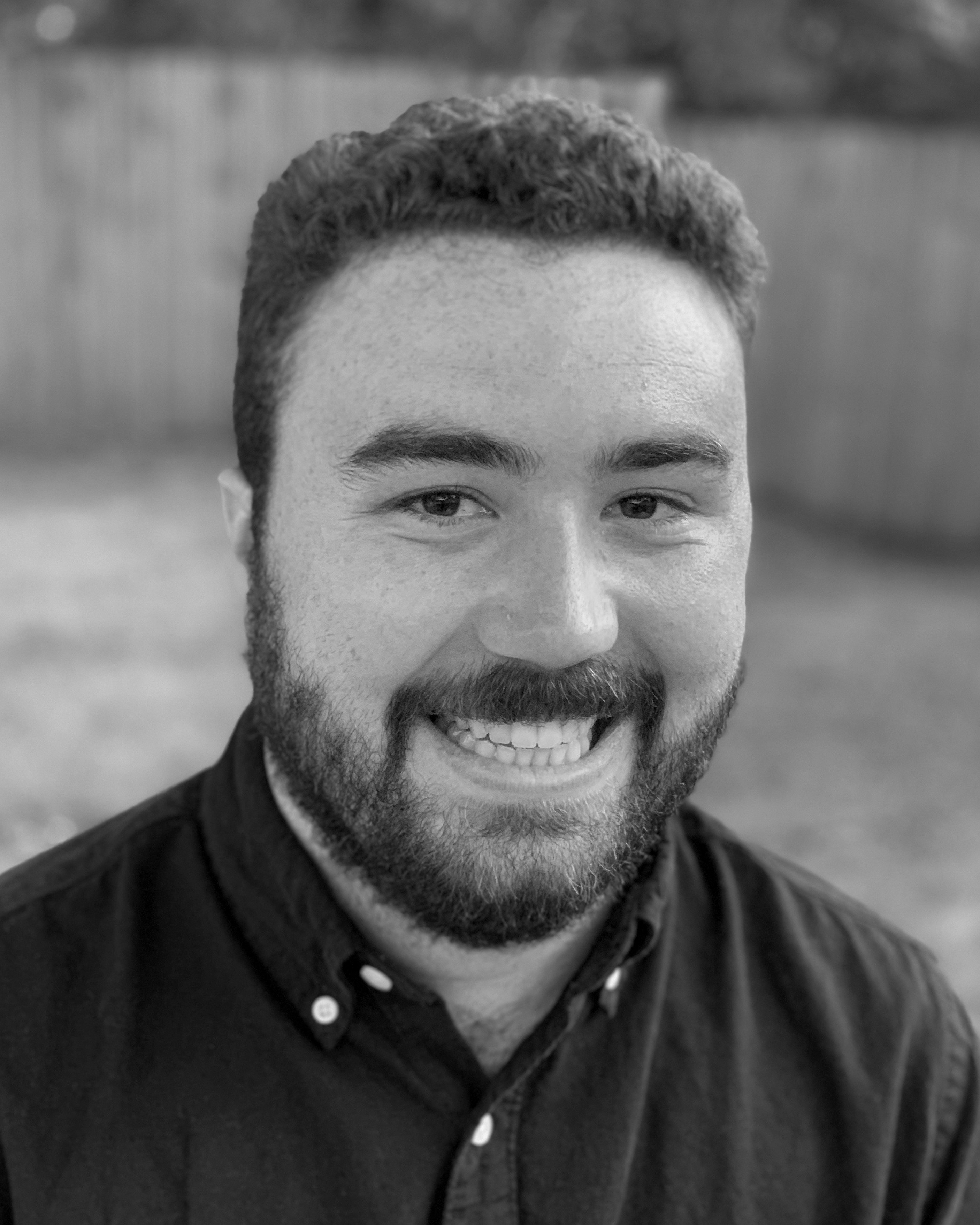}}]{Connor Ligeikis} (S'21) is a Ph.D. Candidate in the Department of Civil and Environmental Engineering at the University of Michigan.  He received his B.S.E. and M.S. degrees in Civil Engineering from the University of Connecticut in 2017 and 2019, respectively, and his M.S. in Electrical and Computer Engineering from the University of Michigan in 2021. His research interests include vibration, control, mechatronics, and cyber-physical systems. He is a National Science Foundation Graduate Research Fellow.
\end{IEEEbiography}

\vspace{-25pt}
\begin{IEEEbiography}[{\includegraphics[width=1in,height=1.25in,clip,keepaspectratio]{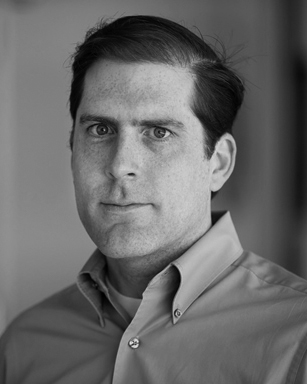}}]{Jeff Scruggs} (M'11) is an Associate Professor in the Departments of Civil and Environmental Engineering, as well as of Electrical and Computer Engineering (by courtesy), at the University of Michigan, which he joined in 2011.  He received his B.S. and M.S. degrees in Electrical Engineering from Virginia Tech in 1997 and 1999, respectively, and his Ph.D. in Applied Mechanics from the Caltech in 2004. Prior to joining the University of Michigan, he held postdoctoral positions at Caltech and the University of California, San Diego, and was on the faculty at Duke University from 2007-11. Scruggs's current research is in the areas of mechanics, vibration, energy, and control.
\end{IEEEbiography}

\end{document}